\documentclass[%
 reprint,
 amsmath,amssymb,
 superscriptaddress,
 aps,
]{revtex4-1}
\usepackage{xcolor}
\usepackage{graphicx}% Include figure files
\usepackage{dcolumn}% Align table columns on decimal point
\usepackage{bm}% bold math
\begin{document}

%\preprint{APS/123-QED}

\title{Efficiency of Local Learning Rules in Threshold-Linear Associative Networks}% Force line breaks with \\

\author{Francesca Sch{\"o}nsberg}
%\email{francesca.schonsberg@sissa.it}
 \affiliation{SISSA, Scuola Internazionale Superiore di Studi Avanzati, Trieste, Italy}%Lines break automatically or can be forced with \\
\author{Yasser Roudi}%
 %\email{yasser.roudi@ntnu.no}
\affiliation{%
 Kavli Institute for Systems Neuroscience \& Centre for Neural Computation, NTNU, Trondheim, Norway\\
}
\author{Alessandro Treves}%
% \email{ale@sissa.it}
 \affiliation{SISSA, Scuola Internazionale Superiore di Studi Avanzati, Trieste, Italy}
 \affiliation{%
 Kavli Institute for Systems Neuroscience \& Centre for Neural Computation, NTNU, Trondheim, Norway\\
}%

\date{\today}% It is always \today, today,

\begin{abstract}
We derive the Gardner storage capacity for associative networks of threshold linear units, and show that with Hebbian learning they can operate closer to such Gardner bound than binary networks, and even surpass it. This is largely achieved through a sparsification of the retrieved patterns, which we analyze for theoretical and empirical distributions of activity. As reaching the optimal capacity via non-local learning rules like backpropagation requires slow and neurally implausible training procedures, our results indicate that one-shot self-organized Hebbian learning can be just as efficient.
\end{abstract}

\pacs{Valid PACS appear here}% PACS, the Physics and Astronomy
\maketitle
\onecolumngrid
\section{Introduction}
Learning in neuronal networks is believed to happen largely through changes in the weights of the synaptic connections between neurons. Local learning rules, those that self-organize through weight changes depending solely on the activity of pre- and postsynaptic neurons, are generally considered to be more biologically plausible than nonlocal ones \cite{brown1990hebbian,*hebb1949organization,*Ami92book}. But how effective are local learning rules? Quite ineffective, has been the received wisdom since the 80's, when nonlocal iterative algorithms came to the fore. However, this wisdom, when it comes to memory storage and retrieval, is largely based on analysing networks of binary neurons \cite{Hop82,Ami+87,Der+87,Gar88}, while neurons in the brain are not binary. 

A better, but still mathematically simple description of neuronal input-output transformation is through threshold-linear (TL) activation function \cite{Har+57,*Har+58,Tre90a}, also predominantly adopted in recent deep learning applications (called ReLu in that context) \cite{Nai+10,*maas2013rectifier,*he2015delving,*goodfellow2016deep}. Therefore, one may ask if the results from the 80's highlighting the contrast between the effective, iterative procedures used in machine learning and the self-organized, one-shot, perhaps computationally ineffective local learning rules are valid beyond binary units \cite{pereira2018attractor}.

The Hopfield model, a most studied model of memory, is a fully connected network of $N$ binary units endowed with a local, {\em Hebbian} learning rule \cite{Hop82,Ami+87}: the weight between two units increases if they have the same activity in a memory pattern; otherwise it decreases. The network can retrieve only up to $p_{max}\simeq 0.14N$ patterns, while, in comparison, Elisabeth Gardner showed \cite{Gar88} that with $C$ connections per unit, the optimal capacity that such a network can attain is $p_{max}=2C$, about 14 times higher; the bound can be approached through iterative procedures like backpropagation that progressively reduce the difference between current and desired output. This consolidated the impression that unsupervised, Hebbian plasticity may well be of biological interest, but is rather inefficient for memory storage. In the fully connected Hopfield model, the transition to no-retrieval is discontinuous: right below the storage capacity, $\sim 1.5\%$ of units in a retrieved pattern are misaligned with the stored pattern, but $50\%$, i.e., chance level, just above the capacity \cite{Ami+87}. This rather low error certainly contributes to the low capacity. However, the negative characterization of Hebbian learning in binary networks persisted even when more errors occur: in the more biologically relevant highly diluted networks the error smoothly goes to $50\%$ \cite{Gar+89}, but the capacity is still a factor of 3 away \cite{Der+87}, {\em approaching} the bound only when the fraction of active unit in each pattern is $f\ll 1$ \cite{Tso+88}.

What about TL units? Are they more efficient in the unsupervised learning of memory patterns? Here we study the optimal pattern capacity \textit{\`a la Gardner} in networks of TL units. Past work discussed above \cite{Tso+88} had suggested that the distribution of activity (along with the connectivity) may play a role in how efficient Hebbian learning is, but, back then, this only meant changing $f$. Besides being a better model of neuronal input-output 
transformation, by allowing non-binary patterns, TL units permit a better understanding of the interplay between the retrieval properties of recurrent networks and the distribution of the activity stored in the network. In fact, we show that while for binary patterns the Gardner bound is larger than the Hebbian capacity no matter how sparse the code, this does not, in general, hold for non-binary stored patterns: the Hebbian capacity can even surpass the bound. This perhaps surprising violation of the bound is because the Gardner calculation imposes an infinite output precision \cite{Bol+93}, while 
Hebbian learning exploits its loose precision to {\em sparsify} the retrieved pattern. In other words, with TL units, Hebbian capacity can get much closer to the optimal capacity or even surpass it, by retrieving a sparser version of the stored pattern. We find that experimentally observed distributions from the Inferior-Temporal (IT) visual cortex \cite{Tre+99}, which can be taken as patterns to be stored, would be sparsified about $50\%$ by Hebbian learning, and would reach about $50\%-80\%$ of the Gardner bound. 

\section{Model Description}
We consider a network of $N$ units and $p$ patterns of activity, $\{\eta_i^{\mu} \}_{i=1,..,N}^{\mu=1,..,p}$ each representing one memory stored in the connection weights via some procedure. Each $\eta_i^{\mu}$ is drawn independently for each unit $i$ and each memory $\mu$ from a common distribution Pr($\eta$). The activity of unit $i$ is denoted by $v_i$ and is determined by the activity of the $C$ units feeding to it as
\begin{subequations}
\begin{align}
&v_i=g[h_i-\vartheta ]^+\\
&h_i\{ v_i \} =\frac{1}{\sqrt{C}}\sum_j J_{ij}v_j,
\end{align}
\label{maineq}
\end{subequations}
where $[x]^+=x$ for $x > 0$ and $=0$ otherwise; and both the gain $g$ and threshold $\vartheta$ are fixed parameters. The \textit{storage capacity}, or capacity for short, is defined as $\alpha_c\equiv p_{\rm max}/C$, with $p_{max}$ the maximal number of memories that can be stored and individually retrieved. The synaptic weights $J_{ij}$ are taken to satisfy the spherical normalization condition for all $i$
\begin{equation}
\sum_{j\neq i}J_{ij}^2=C.
\label{normali_N}
\end{equation}
We are interested in finding the set of $J_{ij}$ that satisfy Eq. \eqref{normali_N}, such that patterns $\{\eta_i^{\mu} \}_{i=1,..,N}^{\mu=1,..,p}$
are self-consistent solutions of Eqs.\ \eqref{maineq}, namely that for all $i$ and $\mu$ we have, $h_i^{\mu}=\vartheta + \eta_i^{\mu}/g$ if  $\eta_i^ {\mu} >0$ and $h_i^{\mu} \le\vartheta$ if $\eta_i^ {\mu} =0$.

\section{Replica Analysis}
Adapting the procedure introduced in \cite{Gar88} for binary units to our network, we evaluate the fractional volume of the space of the interactions $J_{ij}$ which satisfy Eqs. \eqref{maineq}-\eqref{normali_N}, using the replica trick and the replica symmetry ansatz, we obtain the standard order parameters 
$m=\frac{1}{\sqrt{C}}\sum_j J_{ij}$ and $q=\frac{1}{C}\sum_j J_{ij}^a J_{ij}^b$ corresponding, respectively, to the average of the weights within each replica and to their overlap between two replicas $a$ and $b$ (Suppl. Mat. at [URL], Sect. A). Increasing $p$, for $C\to \infty$, shrinks the volume of the compatible weights, eventually to a single point, i.e., when there is only a unique solution and the storage capacity is reached. This corresponds to the case where all the replicated weights are equal $q\to 1$, implying that only one configuration satisfying all the equations 
exists. Adding a further memory pattern would make it impossible, in general, to satisfy them all. At the end, we obtain the following equations for $\alpha_c$
\begin{eqnarray}
0&=&-f(x+\frac{d_1}{g\sqrt{d_3}})+(1-f)\int_x^{\infty}Dt (t-x) \label{final_sol}  \\
\frac{1}{\alpha_c}&=&f \Big [ x^2+\frac{d_2}{g^2 d_3}+\frac{2xd_1
}{g\sqrt{d_3}} +1 \Big ] + (1-f)\int_x^{\infty}Dt(t-x)^2 \nonumber,
\end{eqnarray}
where we have introduced the averages over Pr($\eta$):  $d_1\equiv \langle \eta_i^{\mu} \rangle$, $d_2\equiv \langle (\eta_i^{\mu})^2 \rangle$ and  $d_3\equiv d_2-d_1^2$; $x=(\vartheta-d_1m)/\sqrt{d_3}$ is the normalized difference between the threshold and the mean input, while $f={\rm Pr}(\eta>0)$ is the fraction of active units and $Dt\equiv dt \exp (-t^2/2)/\sqrt{2\pi}$. 
The two equations yield $x$ and $\alpha_c$. Both equations can be understood as averages over units, respectively of the actual input and of the square input, which determine the amount of quenched noise and hence the storage capacity.

%[htb]
\begin{figure}
    \centering
    \begin{tabular}{ll}
    \includegraphics[width=40mm]{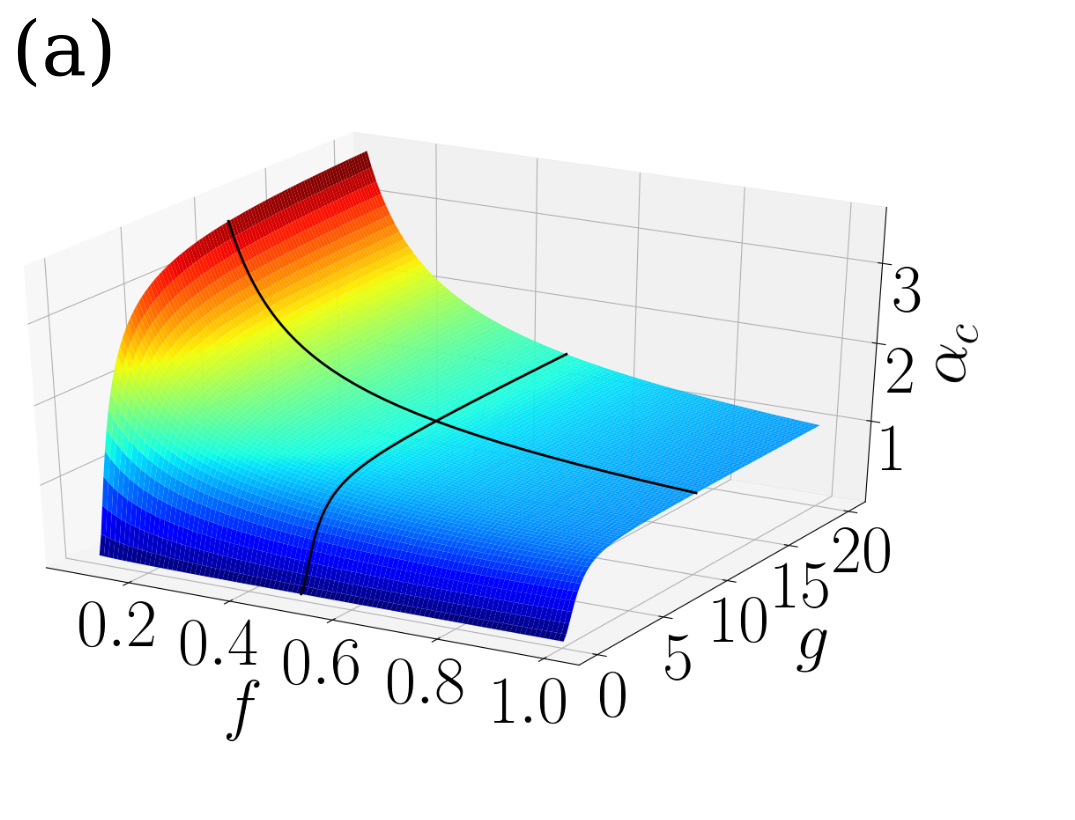}& \includegraphics[width=40mm]{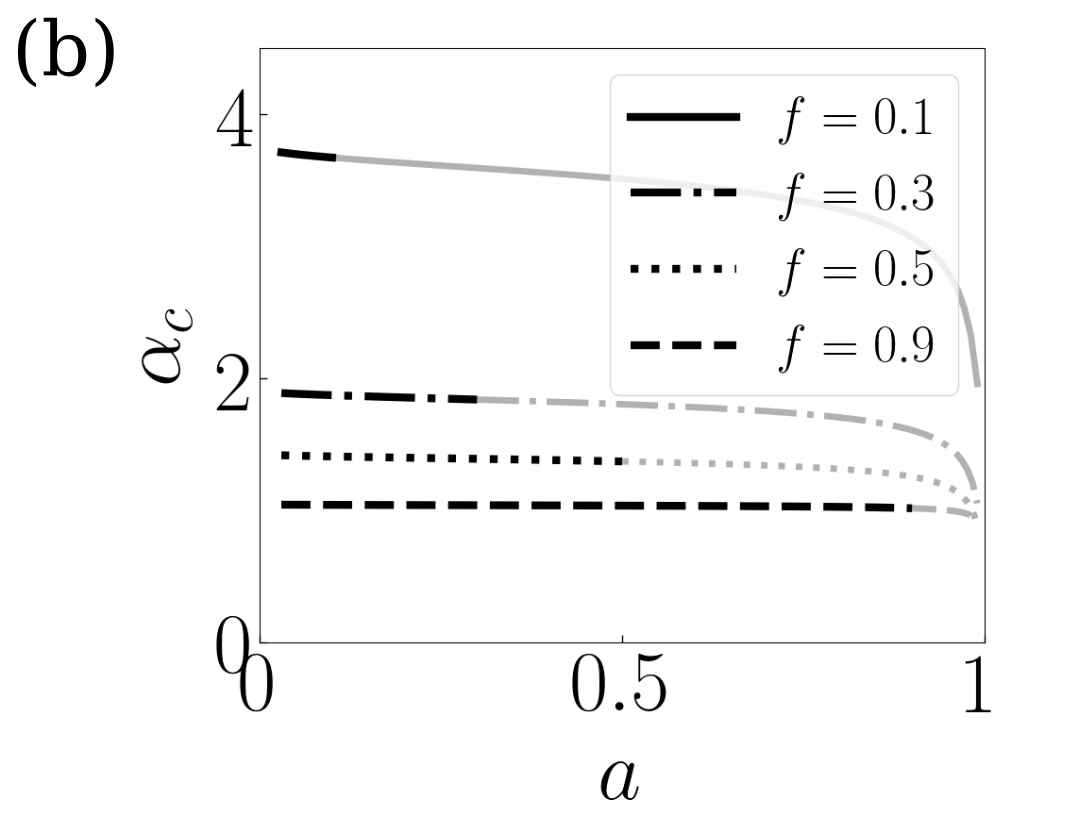} \\
     \includegraphics[width=40mm]{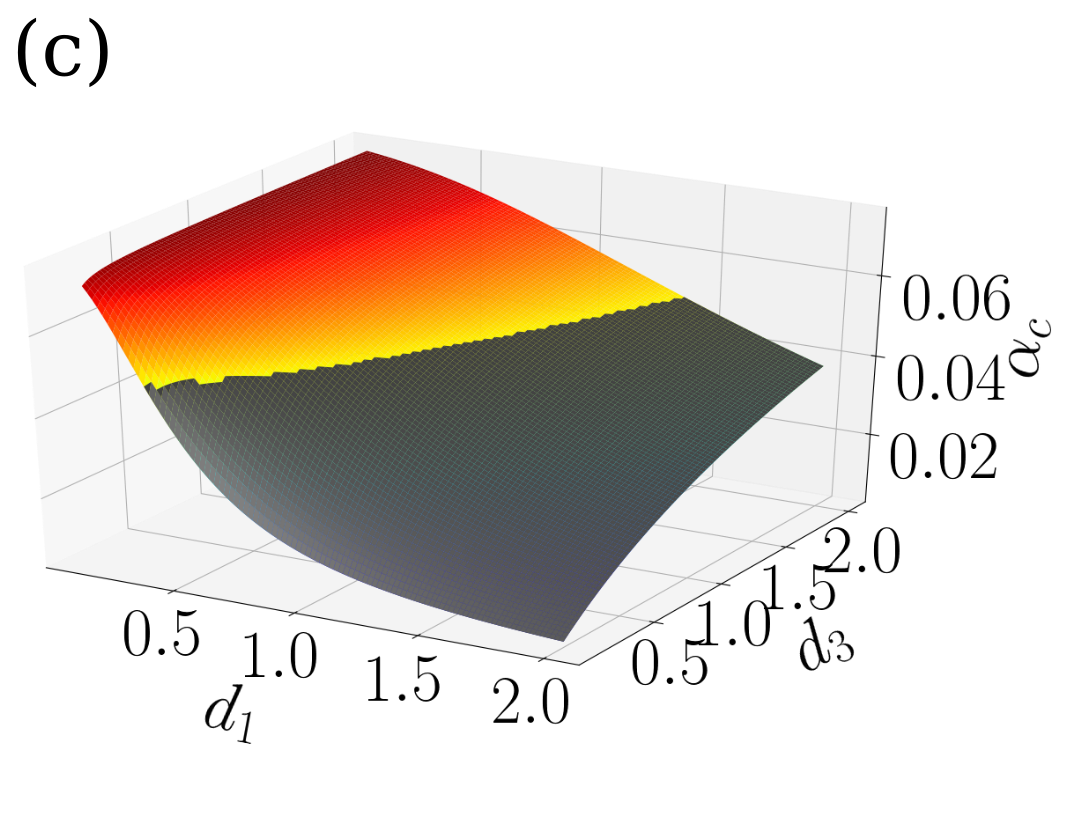}& \includegraphics[width=40mm]{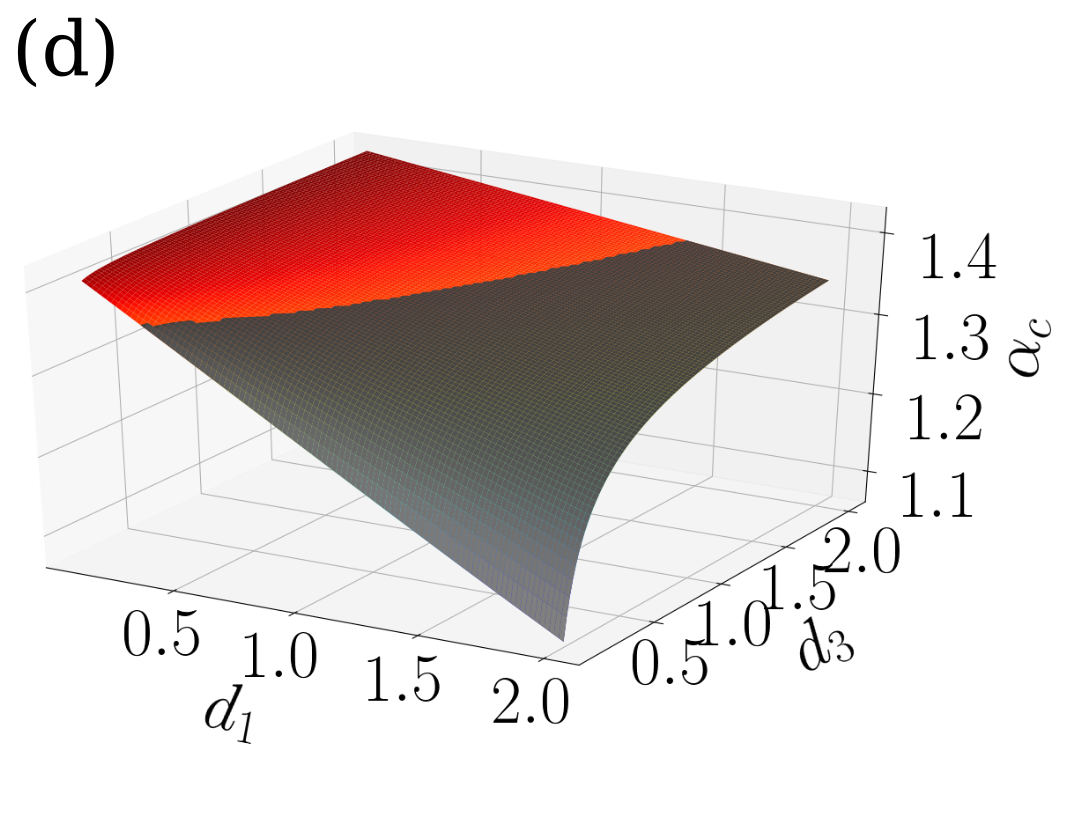}
    \end{tabular}
    \caption{Dependence of the Gardner capacity $\alpha_c$ on different parameters: in (a) as a function of $g$ and $f$ ($d_1=1.1, d_2=2$), in (b)  as a function of $a=d_1^2/d_2$ for different values of $f$ ($g=10$, $d_1=1.1$), in (c) and (d) as a function of $d_1$ and $d_3$ for $g=0.2$ and $g=10$, respectively ($f=0.5$). Note that fixing $f$ restricts the available range of $a$, as $a$ cannot be larger than $f$; the inaccessible ranges are shadowed in (b-d).}
    \label{first_figure}
\end{figure}
The capacity, $\alpha_c$, then depends on the proportion $f$ of active units, but also on the gain $g$, and on the cumulants $d_1$ and $d_3$. Fig.\ \ref{first_figure}a shows that at fixed $g$, $\alpha_c$ increases as more and more units remain below threshold, ceasing to contribute to the quenched noise. In fact, $\alpha_c$ diverges as $[2f\ln(1/\sqrt{2\pi}f)]^{-1}$, for $f\to 0$; see Suppl. Mat. at [URL], Sect. B. At fixed $f$, there is an initially fast increase with $g$ followed by a plateau dependence for larger values of $g$. One can show 
that $\alpha_c\to\frac{g^2}{g^2+1}$ as $f\to 1$, i.e., when all the units in the memory patterns are above threshold, it is always $\alpha_c < 1$ for any finite $g$. 
At first sight this may seem absurd: a linear system of $N^2$ independent equations and $N^2$ variables always has an inverse solution, which would lead to $\alpha_c$ being (at least) one. Similar to what was already noted in \citep{Bol+93}, however, the inverse 
solution does not generally satisfy the spherical constraint in Eq. \eqref{normali_N}; but it does, in our case, in the limit $g\rightarrow \infty$ and this can 
also be understood as the reason why $\alpha_c$ is highest when $g$ is very large. In practice, Fig.\ \ref{first_figure} indicates that over a broad range of $f$ values, $\alpha_c$ approaches its $g\to \infty$ limit already for moderate values of $g$; while the dependence on $d_1$ and $d_3$ is 
only noticeable for small $g$, as can be seen by comparing Fig.\ \ref{first_figure}c and d. For $g\to\infty$, one sees that Eqs. \eqref{final_sol} depend on Pr($\eta$) only through $f$. 

Eqs. \eqref{final_sol}, at $g\to \infty $, have been verified by explicitly training a threshold linear perceptron with binary patterns, 
evaluating $\alpha_c$ numerically as the maximal load which can be retrieved with no errors; See Suppl. Mat. at [URL], Sect. C for details. Estimated values of $\alpha_c$ are depicted by red diamonds in Fig. \ref{Figure_2}, and they follow the profile of the solid line describing the $g\to \infty $ limit of Eq. \eqref{final_sol}.  

\section{Comparison with a Hebbian Rule: theoretical analysis}
With highly diluted connectivity and non-sparse patterns a binary network can get to $1/\pi$ of the bound, even if with vanishing overlaps, much closer than in the fully connected case. This is intuitively because the quenched noise is diminished as $J_{ij}$ and $J_{ji}$ become effectively independent. Besides its biological relevance, with TL units, the fair comparison to the capacity \textit{\`a la Gardner} is thus that of a Hebbian network with {\em highly diluted} connectivity. In what follows, we indicate the Gardner capacity as calculated in the previous section and the Hebbian capacity, by $\alpha^G_c$ and $\alpha_c^H$, respectively, and use similar superscript notations for other quantities.

The capacity of the TL network with diluted connectivity was evaluated analytically in \cite{Tre91,*Tre+91}; see Suppl. Mat. at [URL], Sect. D for a recap, which includes
Refs. \cite{Tre90b, *Rou+06}. Whereas for $g\to\infty $ the Gardner capacity depends on Pr($\eta$) only via $f$, for Hebbian networks it does depend on the distribution, and most importantly on $a$, the {\em sparsity} 
\begin{equation}
a=\langle \eta_i^{\mu} \rangle ^2/\langle (\eta_i^{\mu})^2 \rangle
\end{equation}
whose relation to $f$ depends on the distribution \cite{Tre91, *Tre+91}.

\begin{figure}[htb]
  \centering
    \begin{tabular}{ll}
    \includegraphics[width=43mm]{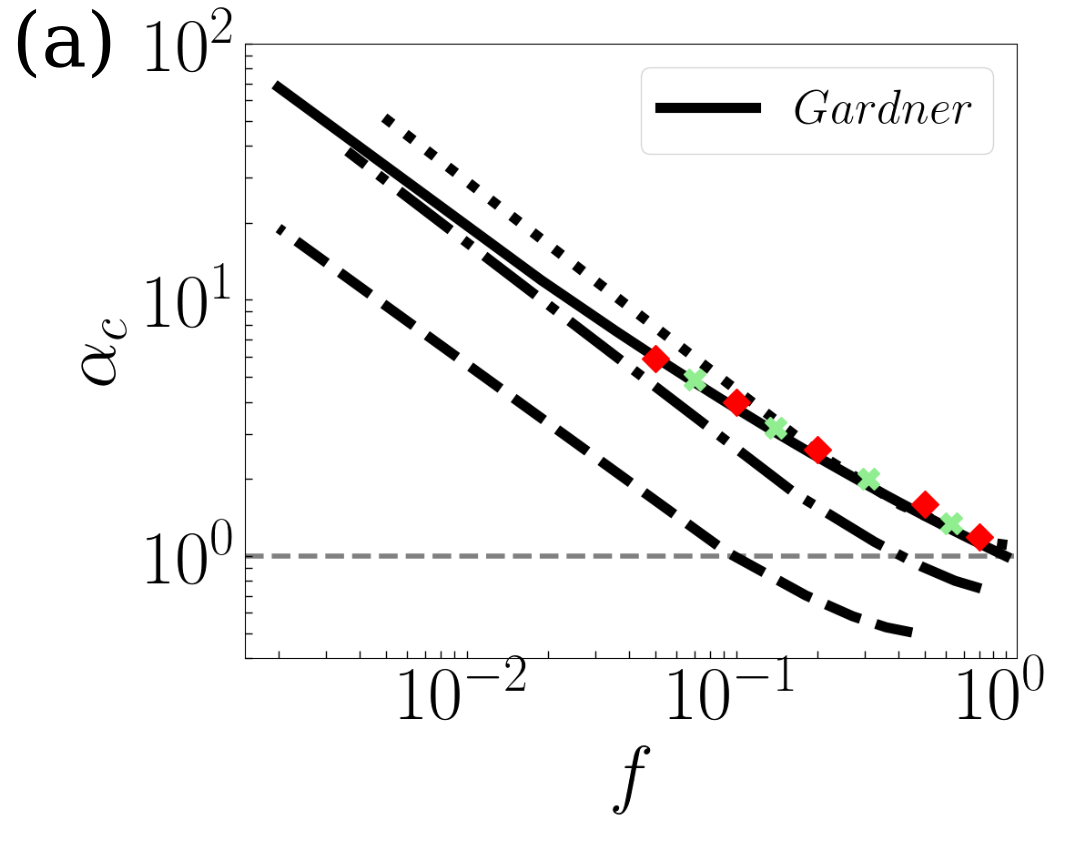}& \includegraphics[width=43mm]{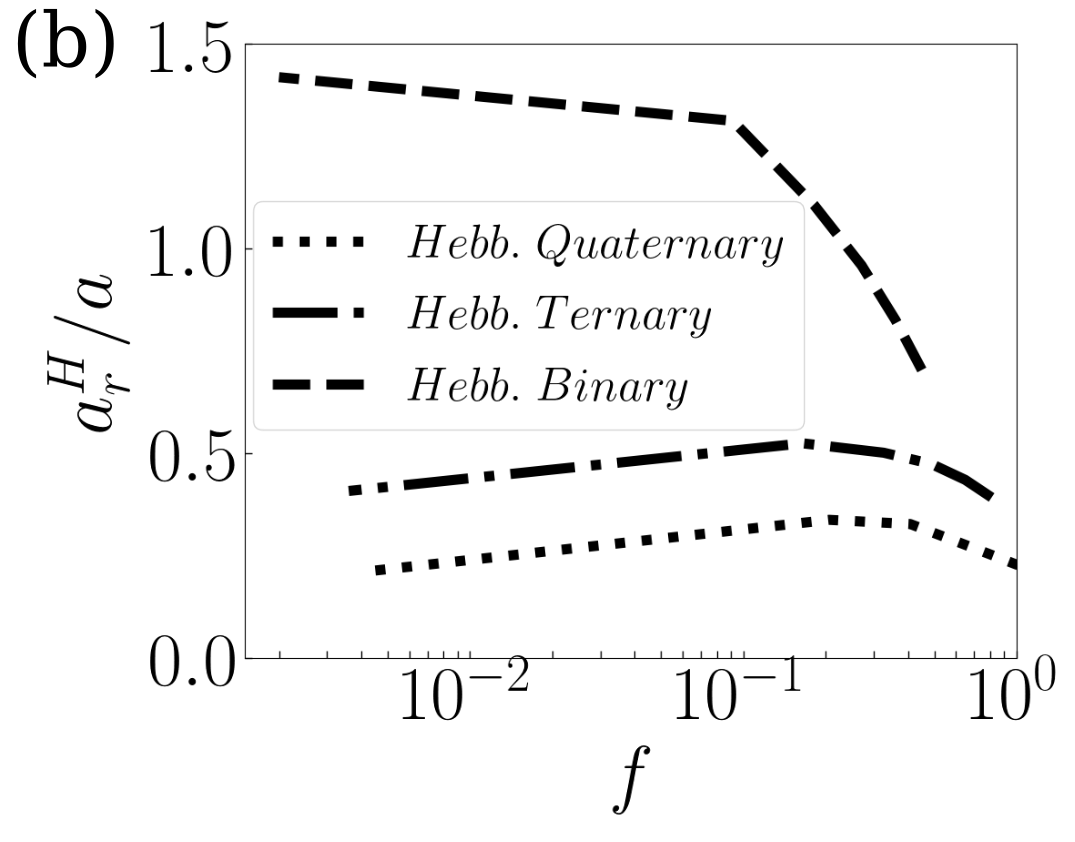}
    \end{tabular}
    \caption{Hebbian vs Gardner capacity. (a) $\alpha_c^H$ vs. $f$ for different sample distribution of stored patterns compared to the analytically calculated universal $\alpha_c^G$; the red diamonds and green crosses are reached using perceptron training for binary and ternary patterns, respectively. (b) the sparsification of the stored patterns at retrieval, for Hebbian networks at their capacity.}
   \label{Figure_2}
\end{figure}

Fig.\ \ref{Figure_2} shows the results for 3 examples of binary, ternary and quaternary distributions for which
 $f$ and $a$ are related through $f=a$, $9a/5$ and $9a/4$, respectively, see Suppl. Mat. at [URL], Sect. E; the Hebbian and the Gardner capacities diverge in the sparse coding limit.

When attention is restricted to binary patterns in Fig.\ \ref{Figure_2}a, the Gardner capacity, $\alpha^G_c$, {\em seems} to provide an upper bound to the capacity reached with Hebbian learning; more structured distributions of activity, however, dispel such a false impression: the quaternary example already shows higher capacity for sufficiently sparse patterns. The bound, in fact, would only apply to perfect errorless retrieval, whereas Hebbian learning 
creates attractors which are, up to the Hebbian capacity limit, correlated but not identical to the stored patterns; in particular, we notice that when considering TL units and Hebbian learning, in order to reach close to the capacity limit, the threshold has to be such as to produce sparser patterns at retrieval, in which only the units with the strongest inputs get activated.
Fig.\ \ref{Figure_2}b shows the ratio of the sparsity of the retrieved pattern produced by Hebbian learning, $a^H_r= \langle v_i^{\mu} \rangle ^2/\langle (v_i^{\mu})^2 \rangle$ (estimated as described in Suppl. Mat. at [URL]) to that of the stored pattern $a$, vs.  $f$: except for the binary patterns at low $f$, the retrieved patterns, at the storage capacity, are always sparser than the stored ones. The largest sparsification happens for quaternary patterns, for which the Hebbian capacity overtakes the Gardner bound, at low $f$. 
Sparser patterns emerge as, to reach close to $\alpha_c^H$, $\vartheta$ has to be such as to inactivate most of the units with intermediate activity levels in the stored pattern. Of course, the perspective is different if $\alpha_c^H$ is considered as a function of $a_r$ instead of $a$, in which case the Gardner 
capacity remains unchanged, as it implies retrieval with $a_r=a$, and is above $\alpha_c^H$ for each of the 3 sample distributions; see Fig. 1 of Suppl. Mat. at [URL].

\section{Comparison with a Hebbian Rule: experimental data}
Having established that the Hebbian capacity of TL networks can surpass 
the Gardner bound for some distributions, we ask what would happen with distributions of firing rates naturally occurring in the 
brain. We considered published distributions of single neurons in IT cortex in response to short naturalistic movies \cite{Tre+99}. Such distributions can be taken as examples of patterns elicited by visual stimuli, to be stored with Hebbian learning, given appropriate conditions, and later retrieved using attractor dynamics, triggered by a partial cue  \cite{fuster1981inferotemporal,*miyashita1988neuronal,*nakamura1995mnemonic,*amit1997paradigmatic,*rolls1998neural,Lim+15}. How many such patterns can be stored, and with what accompanying sparsification?

\begin{figure}[htb]
    \centering
    \includegraphics[scale=0.14]{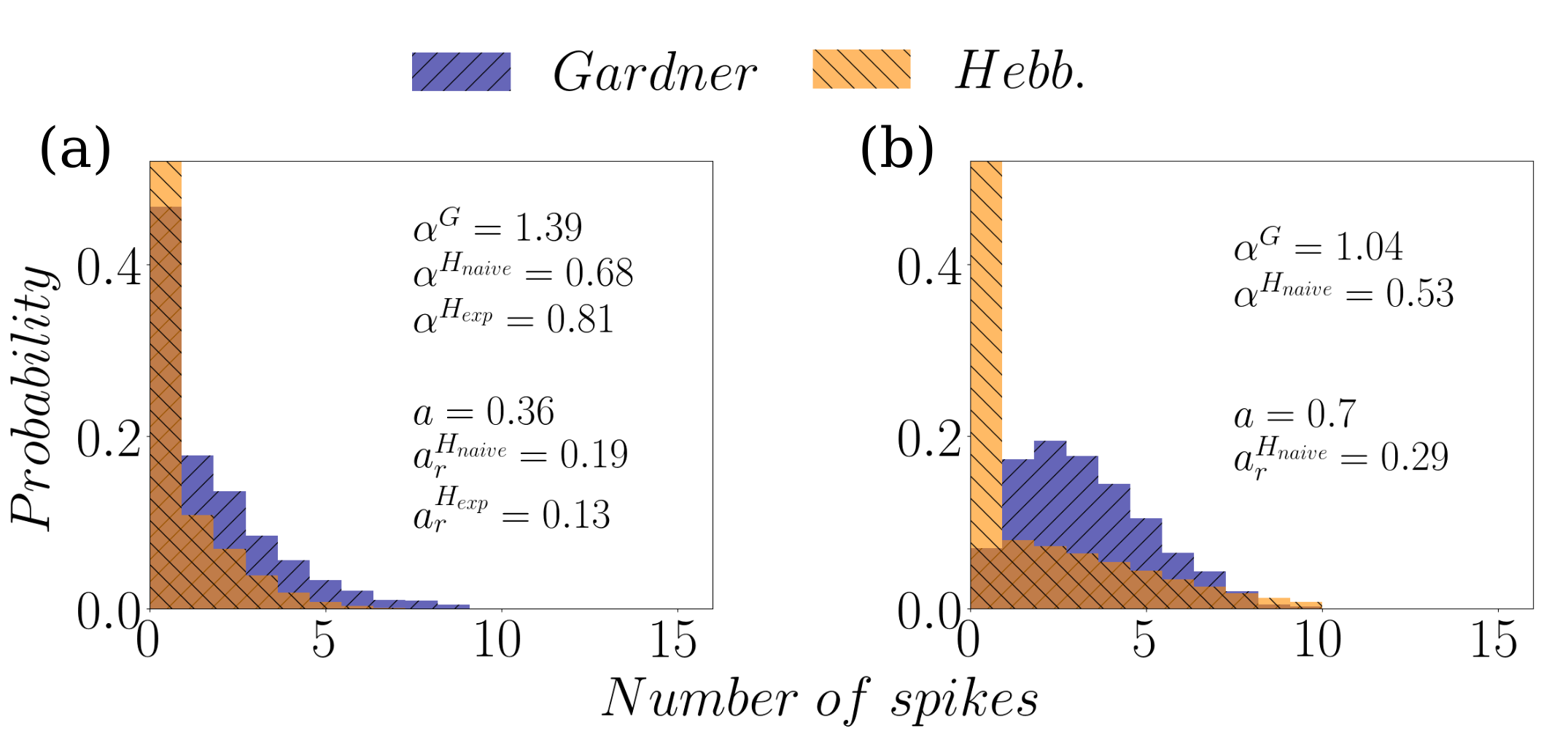}\\
     \includegraphics[scale=0.14]{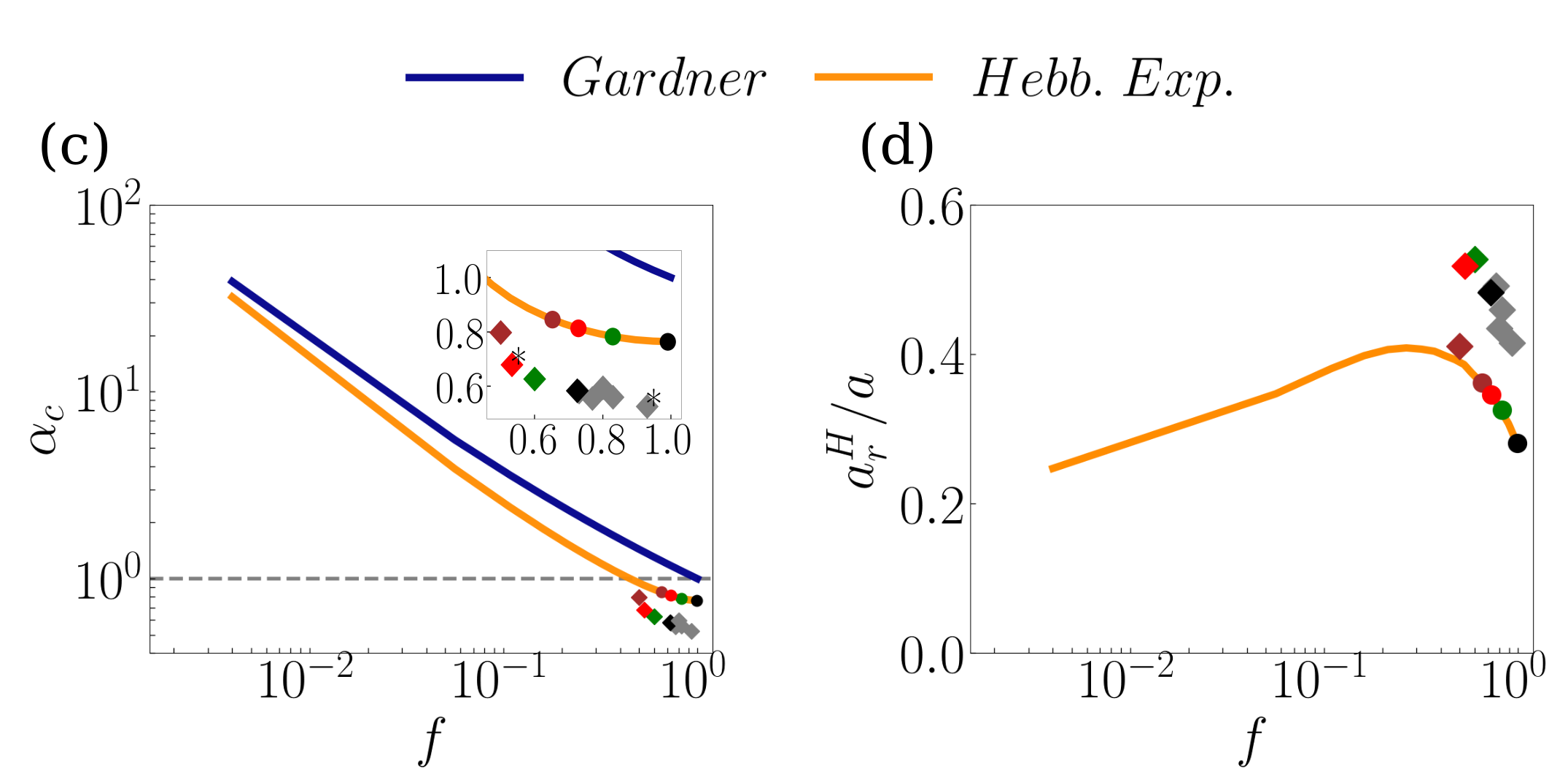}
    \caption{Hebbian vs. Gardner capacity for experimental data. (a,b) histograms of two  experimentally recorded spike counts (blue) and the retrieved distributions, if the patterns were stored using Hebbian learning (orange). Note that the retrieved distributions \textit{\`a la Gardner} would be the same as the stored patterns. (c) Analytically calculated Gardner capacity $\alpha_c^G$ (blue), compared to $\alpha_c^{H_exp}$ for the Hebbian learning of an exponential distribution (orange, circles). $\alpha_c^{H_{naive}}$ is shown by diamonds. The asterisks mark the two cells whose distribution is plotted in (a-b). (d) Sparsification of the retrieved patterns, for Hebbian learning.}
   \label{third_figure}
\end{figure}

Fig.\ \ref{third_figure}a-b show the analysis of two sample distributions from \cite{Tre+99}. The observed distributions, in blue, labeled ``Gardner'', are those we assume could be stored and retrieved, exactly as they were, with a suitable training procedure bound by the Gardner capacity. In orange, we plot the distribution that would be retrieved following Hebbian learning operating at its capacity, see Suppl. Mat. at [URL], Sect. I for the estimation of the retrieved distribution. Note that the absolute scale of the retrieved firing rate is arbitrary, what is fixed is only the shape of the distribution, which is sparser (as clear already from the higher bar at zero). The pattern in Fig.\ \ref{third_figure}a, which has $a<0.5$, could also be fitted with an exponential distribution having $f=2a$ (see Suppl. Mat. at [URL], Sect. F). 
In that panel we also show the values of $\alpha_c^{H_{exp}}$ and $a^{H_{exp}}_r$, calculated assuming the exponential fit, along with values from the observed discrete distribution ($\alpha_c^{H_{naive}}$ and $a^{H_{naive}}_r$).
Fig.\ \ref{third_figure}c shows both $\alpha_c^G$ and $\alpha_c^{H_{exp}}$ versus $f$; we have indicated by diamonds the Hebbian capacities for the $9$ empirical distributions in \cite{Tre+99} and by circles the fitted values for those which could be fitted to an exponential. In the  Suppl. Mat. at [URL], Sect. G we also discuss the fit to a log-normal, which is better at reproducing experimental distributions with a mode above zero \cite{buzsaki2014log}, as in Fig.\ \ref{third_figure}b.

There are three conclusions that we can draw from these data. First, the Hebbian capacity from the empirical distributions is about $80 \%$ of that of the exponential fit, when available.  Second, in general for distributions like those of these neurons, the capacity achieved by Hebbian learning is about 
$50 \%-80\%$ of the Gardner capacity, depending on the neuron and whether we take its discrete distribution \textit{as is}, or fit it to an exponential (or, e.g., to a log-normal) shape. 
Third, with Hebbian learning retrieved patterns tend to be $2-3$ times sparser than the stored ones, again depending on the particular distribution, empirical or exponential fit (as for non-sparse distributions, which could be better fit by a log-normal, see Suppl. Mat. at [URL], Sect. G). 
As illustrated in  Fig.\ \ref{third_figure}d, the empirical distributions achieve a lower capacity than that of their exponential fit, as the latter leads to further sparsification at retrieval.

\section{Discussion}--.
While instrumental in conceptualizing memory storage \cite{amit1997dynamics,*Bat+98,*Yoo+13}, Hebbian learning has been widely considered a poor man's option, relative to more powerful machine learning algorithms that could reach the Gardner bound for binary units and patterns. No binary or quasibinary pattern of activity has ever been observed in the cerebral cortex, however. A few 
studies have considered TL units, showing them to be less susceptible to memory mix-up effects \cite{Tre91b,*Rou+03} or perturbations in the weights and inputs values \cite{Bal+19} but, in the framework of \textit{\`a la Gardner} calculations, they have focused on issues other than associative networks storing sparse representations. 
For instance, a replica analysis was carried out in \cite{Bol+93} with a generic gain function, but then discussed only in a quasibinary regime. Others considered monotonically increasing activation functions under the constraint of nonnegative weights \cite{Clo+13}. Here, we report the analytical derivation of the Gardner capacity for TL networks, validate it via perceptron training, and compare it with Hebbian learning. We find that the bound can be reached or even surpassed, and that retrieval leads to sparsification. For sample experimental distributions, we find that one-shot Hebbian learning can utilize $50-80\%$ of the available ``errorless'' capacity if retrieving sparser activity, compatible with recent observations \cite{Lim+15}.

In deriving the Gardner bound, we assumed errorless retrieval and it remains to be seen how much allowing errors increases this bound for TL units and neurally plausible distributions. For the binary case of \cite{Gar+89}, as already mentioned, this errorless bound is still above the Hebbian capacity of the highly diluted regime, with its continuous (second order) transition, i.e., with vanishing overlap at storage capacity \cite{Gar+89}.
How does the overlap behave in the TL case? For highly diluted TL networks with Hebbian learning, in fact, except for special cases, the transition at capacity is discontinuous: the overlap drops to zero from a non-zero value that depends on the distribution of stored neural activity but can be small \cite{Sch+21}.
It is worth noting, though, that while in the binary case the natural measure of error is simply the fraction of units misaligned at retrieval, in the TL case error can be quantified in other ways. In the extreme in which only the most active cells remain active at retrieval, those retrieved memories cannot be regarded as the full pattern, with its entire information content, but more as a pointer, effective perhaps as a mechanism only to distinguish between different possible patterns or to address the full memory elsewhere, as posited in {\em index} theories of 2-stage memory retrieval \cite{Tey+86}.
Further understanding would also derive from comparing the maximal information content per synapse for TL units, with Hebbian or iterative learning, as previously studied for binary networks \cite{Nad+90}. Using nonbinary patterns might also afford a solution to the low storage capacity observed in balanced memory networks storing binary patterns \cite{roudi2007balanced}.

Our focus here has been on memory storage in associative neural networks, with the overarching conclusion that the relative efficiency of Hebbian learning is much higher when units have a similar transfer function to real cortical neurons. The efficiency of local learning rules had also been challenged by their comparatively weaker performance in other (machine learning) settings \cite{NIPS2018_8148}, while results to the contrary are also reported \cite{amit2019deep,krotov2019unsupervised}. It may therefore be argued that the efficiency of local learning in these settings might also be fundamentally dependent on both the types of units used and the data, observations consistent with the findings in \cite{krotov2019unsupervised} and \cite{NIPS2018_8148}, respectively. In evaluating a learning rule, it may therefore be crucial to consider whether it is suited to the transfer function and data representation it operates on. 

\begin{acknowledgments}
We thank R\'emi Monasson and Aldo Battista for useful comments and discussions. This research received funding from the EU Marie Skłodowska-Curie Training Network 765549 ``M-Gate'', and partial support from Human Frontier Science Program RGP0057/2016, the Research Council of Norway (Centre for Neural Computation, grant number 223262; NORBRAIN1, grant number 197467), and the Kavli Foundation.\\

Y.R and A.T. contributed equally to this work.
\end{acknowledgments}
\bibliography{mybibAPS}{}
\end{document}

\onecolumngrid
\section*{Supplemental Material}
%======================================================
\subsection{Derivation of the Gardner bound for TL units}
In this section we give a detailed mathematical derivation of the Gardner bound reported in the main text. 

We start by considering a single threshold-linear unit whose activity is denoted by $u$. The unit receives $C$ inputs $v_j$, for $j=1\cdots C$, through synaptic weights $J_j$. The activity of the unit is determined through the threshold-linear activation function as
\begin{eqnarray}
&&u=g[h_i-\vartheta ]^+\nonumber\\
&&h\{ v \} =\frac{1}{\sqrt{C}}\sum_j J_{j}v_j,
\end{eqnarray}
We assume that we have $p$ patterns of activity over the inputs, that we denote by $\xi^{\mu}_j$, with $\mu=1\cdots p $. For each input pattern $\mu$ we also consider a desired output activity for each unit that we denote $\eta^\mu$. We are interested in finding how many patterns can be stored in the synaptic weights, such that the input activity elicits the desired output activity, assuming that the synaptic weights satisfy the spherical constraint
\begin{equation}
\sum_{j\neq i}J_{j}^2=C.
\label{spherical}
\end{equation}
Following \cite{Gar88}, the fractional volume in the space of interactions $J$ that satisfy Eq.\ \eqref{spherical} and the correct output  $\eta^{\mu}$ given the inputs $\xi^{\mu}_j$ can be written as
\begin{equation}
V=\frac{ \int \prod_{j,j\neq i} d J_{j} \delta\left (\sum_{j} J^2_{j} - C\right)\prod_{\mu}\Big[  \left(1-\delta_{\eta^{\mu},0}\right)\delta\left(h^{\mu} - \vartheta-\frac{\eta^{\mu}}{g}\right)+\delta_{\eta^{\mu},0}\Theta\left(\vartheta-h_i^{\mu}\right) \Big] }{\int \prod_{j, j\neq i} d J_{j} \delta \left(\sum_{j} J^2_{j} - C\right)},
\end{equation}
Calculating the optimal capacity essentially boils down to calculating, in the thermodynamic limit $C\rightarrow \infty$, the expectation of the logarithm of this fractional volume $V$ over the distribution of $\eta$ and $\xi$ and finding for what value of $p$ it shrinks to zero. For calculating $\langle \text{ln} V \rangle_{\eta,\xi}$, we use the replica trick $\langle \ln V \rangle=\lim_{n\to 0}\frac{\langle V^n \rangle - 1}{n}$, which turns the problem to that of computing the replica average $\langle V^n \rangle_{\xi,\eta}$, namely 
\begin{equation}
\langle V^n \rangle_{\xi,\eta}=\Bigg\langle
\prod_{a=1,..,n}\prod_{\mu} 
\frac{ \int \prod_{j, j\neq i} d J^a_{j} \delta\left(\sum_{j} (J_{j}^a)^2 - C\right)\Big[ 
\left (1-\delta_{\eta^{\mu},0}\right )\delta \left(h^{a,\mu} -  \vartheta-\frac{\eta^{\mu}}{g}\right)+\delta_{\eta^{\mu},0}\Theta( \vartheta-h^{a,\mu})\Big]  }{\int \prod_{j,j \neq i} d J^a_{j}  \delta\left(\sum_{j} (J_{j}^a)^2  - C\right)}\Bigg\rangle_{\xi, \eta}.
\label{logV}
\end{equation}
We first compute the numerator. To compute the averages over $\xi$  and $\eta$ in the numerator, we note that the delta function can be written as
\begin{equation}
\begin{split}
\delta(h^{a,\mu} - \vartheta-\frac{\eta^{\mu}}{g}) &= \int\frac{dx_{\mu}^a}{2\pi}\exp\Bigg \{ ix_{\mu}^a \Big ( \frac{1}{\sqrt{C}}\sum_{j}J_{j}^a \xi^{\mu} - \vartheta -\frac{\eta^{\mu}}{g}    \Big )   \Bigg \} \\
&=\int\frac{dx_{\mu}^a}{2\pi}
\exp \Big [ - \frac{i x_{\mu}^a}{g}\Big ( \eta^\mu + g\vartheta     \Big )   \Big ] 
\exp \Big [ \frac{i x_{\mu}^a \sum_{j} J_{j}^a\xi^{\mu}}{\sqrt{C}}  \Big ].
\end{split}
\label{identity_delta}
\end{equation}
For the average of the Heaviside function, we write
\begin{equation}
\begin{split}
\Theta(\vartheta-h^{a,\mu}) &= \int_0^{\infty} d\lambda_{\mu}^a\delta[\lambda_{\mu}^a-(\vartheta-h^{a,\mu})]\\
&= \int_0^{\infty} \frac{d\lambda_{\mu}^a}{2\pi}\int_{-\infty}^{\infty}dy_{\mu}^a\exp[i y_{\mu}^a(\lambda_{\mu}^a-(\vartheta-h^{a,\mu}))]\\
&=  \int_0^{\infty} \frac{d\lambda_{\mu}^a}{2\pi}\int_{-\infty}^{\infty}dy_{\mu}^a
\exp \Big [i y_{\mu}^a ( \lambda_{\mu}^a-\vartheta)  \Big ]
\exp \Big [ \frac{i y_{\mu}^a \sum_{j} J_{j}^a\xi^{\mu}}{\sqrt{C}}  \Big ].
\end{split}
\label{identity_theta}
\end{equation}
We now use the above identities in Eqs.\ \eqref{identity_delta} and \eqref{identity_theta} to compute the following quantity that appears in the numerator of Eq.\ \eqref{logV}, assuming independently drawn $\xi$ and $\eta$ as
\begin{equation}
\begin{split}
e^{CM} &\equiv\Bigg \langle  \prod_{\mu,a} (1-\delta_{\eta^{\mu},0})\delta(h^{a,\mu} - \vartheta-\frac{\eta^{\mu}}{g})+\delta_{\eta^{\mu},0}\Theta(\vartheta-h^{a,\mu}) \Bigg \rangle_{\xi,\eta} \\
&=\prod_{\mu} \Bigg \langle (1-\delta_{\eta^{\mu},0}) \Big \langle  \prod_a \delta(h^{a,\mu} - \vartheta-\frac{\eta^{\mu}}{g}) \Big \rangle_{\xi^{\mu}} + \delta_{\eta^{\mu},0} \Big \langle  \prod_a \Theta(\vartheta-h^{a,\mu})  \Big \rangle_{\xi^{\mu}}   \Bigg \rangle _{\eta^{\mu}}.
\end{split}
\label{eCM}
\end{equation}
In order to compute the average of the delta functions in Eq.\eqref{eCM},  we use the approximation
\begin{eqnarray*}
\left \langle \exp(x) \right \rangle&=&\langle1+x+\frac{x^2}{2}+\mathcal{O}(x^3)\rangle=1+\langle x \rangle+\frac{\langle x^2 \rangle}{2}+\langle \mathcal{O}(x^3) \rangle\\
&\approx& \exp\left \{ \langle x \rangle +\frac{ \langle x^2 \rangle}{2}-\frac{ \langle x \rangle^2}{2}\right\}
\end{eqnarray*}
to calculate the following average
\begin{equation}
\begin{split}
&\left\langle  \exp \left \{ \frac{i\sum_{a,j} x^a_{\mu} J^a_{j} \xi^{\mu}_j}{\sqrt{C}}\right \}  \right \rangle_{\xi^{\mu}} =\\  
&= \exp\left \{\frac{i}{\sqrt{C}} \sum_{a,j} x^{a}_{\mu}J^{a}_{j} \langle \xi^{\mu}_j \rangle 
-\frac{1}{2C}\sum_{a,b,j,k} x^{a}_{\mu} x^{b}_{\mu}J^{a}_{j} J^{b}_{k} \langle \xi^{\mu}_j  \xi^{\mu}_k \rangle 
-\frac{1}{2} \left (\frac{i}{\sqrt{C}} \sum_{a,j} x^{a}_{\mu} J^{a}_{j}  \langle \xi^{\mu}_j  \rangle \right)  \left (\frac{i}{\sqrt{C}} \sum_{b,k} x^{b}_{\mu} J^{b}_{j}  \langle \xi^{\mu}_k  \rangle \right) \right\} \\ 
&=\exp \Bigg \{ \frac{i}{\sqrt{C}} \sum_{a,j} x^{a}_{\mu}J^{a}_{j} \langle \xi^{\mu}_j \rangle -\frac{1}{2C} \Big [ \sum_{a,b,j} x^{a}_{\mu} x^{b}_{\mu}J^{a}_{j} J^{b}_{j} \langle (\xi^{\mu}_j)^2  \rangle + \sum_{a,b,j,k!=j} x^{a}_{\mu} x^{b}_{\mu}J^{a}_{j} J^{b}_{k} \langle \xi^{\mu}_j \rangle \langle \xi^{\mu}_k \rangle \Big ] +\\
& + \frac{1}{2C} \Big [ (\sum_{a,b,j} x^{a}_{\mu} x^{b}_{\mu} J^{a}_{j}J^{b}_{j}\langle \xi^{\mu}_j  \rangle^2 ) +\sum_{a,b,j,k!=j} x^{a}_{\mu} x^{b}_{\mu} J^{a}_{j}J^{b}_{k}  \langle \xi^{\mu}_j  \rangle\langle \xi^{\mu}_k  \rangle \Big ] \Bigg \} \\
&= \exp\left \{\frac{i}{\sqrt{C}} \sum_{a,j} x^{a}_{\mu}J^{a}_{j} \langle \xi^{\mu}_j  \rangle -\frac{1}{2C}\sum_{a,b,j} x^{a}_{\mu} x^{b}_{\mu}J^{a}_{j} J^{b}_{j} \langle (\xi^{\mu}_j)^2\rangle +\frac{1}{2C}\sum_{a,b,j} x^{a}_{\mu} x^{b}_{\mu}J^{a}_{j} J^{b}_{j} \langle \xi^{\mu}_j  \rangle^2 \right\} 
\end{split}
\label{expexpand}
\end{equation}
where in going from the second to the third line in Eq.\ \eqref{expexpand}, we have used the fact that $\langle \xi^{\mu}_j \xi^{\mu}_k \rangle = \langle \xi^{\mu}_j \rangle \langle \xi^{\mu}_k\rangle$.
Expanding the second exponential in the second line of Eq. \eqref{identity_delta}, we can write, in the large $C$ limit
\begin{equation}
\begin{split}
&\Big \langle  \prod_a \delta(h^{a,\mu} - \vartheta-\frac{\eta^{\mu}}{g}) \Big \rangle_{\xi^{\mu}} =\\  
&= \int_{-\infty}^{\infty} \Big [ \prod_a \frac{d x_{\mu}^a}{2\pi}\Big ]
\exp \Big [ -\frac{i}{g}(\eta^{\mu} + g \vartheta)\sum_a x_{\mu}^a + i d_1^{inp} \sum_a x_{\mu}^a m^a -\frac{d_3^{inp}}{2} \Big ( \sum_a (x_{\mu}^a)^2 + 2\sum_{a<b}x_{\mu}^a x_{\mu}^b q^{ab} \Big ) \Big ]\\
&\equiv I_1(q^{ab},m^a,\eta^{\mu})
\end{split}
\label{i1}
\end{equation}
in which we have assumed symmetric replicas and defined $d_1^{inp}\equiv \langle \xi_j^\mu \rangle$, $d_2^{inp}\equiv \langle (\xi_j^\mu )^2\rangle$, $d_3^{inp}\equiv d_2^{inp}- (d_1^{inp})^2$ and 
\begin{subequations}
\begin{eqnarray}
&q^{ab}=\frac{1}{C}\sum_j J^a_{j} J^{b}_{j}\\
&m^{a}=\frac{1}{\sqrt{C}}\sum_j J^a_{j}
\end{eqnarray}
\label{qmdef}
\end{subequations}
Similarly, using the identity in Eq. \eqref{identity_theta} we have
\begin{equation}
\begin{split}
&\Big \langle  \prod_a \Theta(\vartheta-h^{a,\mu})  \Big \rangle_{\xi^{\mu}}= \\
&=\int_0^{\infty}\Big [  \prod_a \frac{d \lambda_\mu^a}{2\pi}\Big ]\int_{-\infty}^{\infty} \Big [ \prod_a d y_{\mu}^a\Big ]
\exp \Big [ i\sum_a(\lambda_{\mu}^a-\vartheta)y_{\mu}^a + id_1^{inp}\sum_a y_{\mu}^a m^a -\frac{d_3^{inp}}{2} \Big ( \sum_a (y_{\mu}^a)^2 +2 \sum_{a<b} y_{\mu}^a y_{\mu}^b q^{ab}   \Big) \Big ]\\
&\equiv I_2(q^{ab}, m^a).
\end{split}
\label{i2}
\end{equation}
Using Eq. \eqref{i1} and \eqref{i2}, the quantity $M(q^{ab},m^a)$ defined through Eq. \eqref{eCM} can be written as
\begin{equation}
M(q^{ab},m^a)=\frac{p}{C}\ln \left [ \langle (1-\delta_{\eta^{\mu},0})I_1(q^{ab},m^a,\eta^{\mu}) +    \delta_{\eta^{\mu},0}I_2(q^{ab}, m^a) \rangle_{\eta^{\mu}} \right ].
\label{M}
\end{equation}
We now insert Eq. \eqref{M} back to Eq. \eqref{logV} and enforce the definitions of $m$ and $q$ in Eq.\ \eqref{qmdef} using the identities
\begin{equation}
\begin{split}
&1=C\int\frac{dq^{ab}d\hat{q}^{ab}}{2i\pi}\exp \left(-C\hat{q}^{ab}q^{ab}+\hat{q}^{ab}\sum_j J_{j}^aJ_{j}^b \right) \\
&1=\sqrt{C}\int \frac{dm^a d\hat{m}^a}{2i\pi}\exp \left(-\sqrt{C}\hat{m}^a m^a +\hat{m}^a\sum_j J_{j}^a \right)
\end{split}
\end{equation}
and the normalization of Eq. \eqref{logV} using
\begin{equation}
\delta \left (\sum_{j} J^{a^2}_{j} - C\right)=\int \frac{d E^a}{4i\pi} \exp \Big(-\frac{E^a}{2} \sum_{j\neq i}J_{j}^{a^2} +\frac{C E^a}{2} \Big)
\end{equation}
such that the numerator in Eq. \eqref{logV} can be written as 
\begin{equation}
\begin{split}
&A=\int 
\Big [ 	\prod_a	\frac{d E^a}{4i\pi}	\Big ]
\Big [ 	\prod_a	\sqrt{C} \frac{dm^a d\hat{m}^a}{2i\pi}	\Big ]
\Big [ 	\prod_{a<b}	C \frac{dq^{ab}d\hat{q}^{ab}}{2i\pi}	\Big ]\\
&e^{C[M(q,m) -\frac{1}{\sqrt{C}}\sum_a \hat{m}^a m^a - \sum_{a<b} \hat{q}^{ab}q^{ab}+\sum_a\frac{E^a}{2}]} 
\int \Big[\prod_{j,a}dJ_{ij}^a\Big]e^{-\sum_{a,j}\frac{E^a}{2}(J_{j}^a)^2+\sum_{a,j}\hat{m}^aJ_{j}^a+\sum_{a<b}\hat{q}^{ab}J_{ij}^aJ_{ij}^b}.
\end{split}
\end{equation}
Defining the function
\begin{equation}
W(\hat{q}^{ab},\hat{m}^a,E^a)=\ln \int \Big [\prod_a d J^a \Big ]\exp 
\Big ( 
-\frac{1}{2}  \sum_a E^a(J^a)^2 + \sum_a\hat{m}^aJ^a+\sum_{a<b}\hat{q}^{ab}J^aJ^b
\Big )
\label{W}
\end{equation}
we can write
\begin{equation}
A=\int 
\Big [ 	\prod_a	\frac{d E^a}{4i\pi}	\Big ]
\Big [ 	\prod_a	\sqrt{C} \frac{dm^a d\hat{m}^a}{2i\pi}	\Big ]
\Big [ 	\prod_{a<b}	C \frac{dq^{ab}d\hat{q}^{ab}}{2i\pi}	\Big ]
e^{C[M(q^{ab},m^{a}) + W(\hat{q}^{ab},\hat{m}^a,E^a) -\frac{1}{\sqrt{C}}\sum_a \hat{m}^a m^a - \sum_{a<b} \hat{q}^{ab}q^{ab}+\sum_a\frac{E^a}{2}]} 
\label{A}
\end{equation}
We can then compute $A$ in Eq. \eqref{A} using the saddle point approximation, by maximizing the argument of the exponential, that is maximising 
\begin{equation}
G(q^{ab},\hat{q}^{ab},m^a,\hat{m}^a,E^a) \equiv M(q^{ab},m^a)+W(\hat{q}^{ab},\hat{m}^a,E^a)-\frac{1}{\sqrt{C}}\sum_a \hat{m}^a m^a - \sum_{a<b} \hat{q}^{ab}q^{ab}+\sum_a\frac{E^a}{2}. 
\end{equation} 
In order to proceed to make this extremisation we assume a replica symmetric ansatz:
\begin{equation}
\begin{split}
& q^{ab}=q	\\
& \hat{q}^{ab}=\hat{q} 	\\
& m^a = m 	\\
& \hat{m}^a =\hat{m}	\\
& E_a= E
\end{split}
\end{equation} 
with these assumptions
\begin{equation}
G(q,\hat{q},m,\hat{m},E)= M(q,m)+W(\hat{q},\hat{m},E) +\frac{n}{2}(-\frac{2\hat{m}m}{\sqrt{C}}+\hat{q}q+E).
\label{g_rsa}
\end{equation}
In the above Eq. \eqref{g_rsa},  $W$ and $M$ are calculated using the limits for $n\rightarrow 0$ of the expressions in Eq. \eqref{M} and \eqref{W}, as follows. For $W$, we use the Gaussian trick (i.e. the one dimensional Hubbard-Stratonovich transformation) 
\begin{equation}
\begin{gathered}
\int_{-\infty}^{\infty} dt \exp (-at^2\pm bt)=\exp \Big [ \frac{b^2}{4a}\Big ]\sqrt{\frac{\pi}{a}}\\
\to e^{-x^2/2}=\int_{-\infty}^{\infty} \frac{dt}{\sqrt{2\pi}} e^{-t^2/2\pm tx} 
\end{gathered}
\label{Gaussian}
\end{equation}
combined with the replica symmetric expression for $W$ to get
\begin{equation}
\begin{split}
W(\hat{m},\hat{q},E) &=\ln \int \Big [\prod_a d J^a \Big ]\exp 
\Big ( 
-\frac{E}{2}  \sum_a (J^a)^2 + \hat{m}\sum_aJ^a+\frac{\hat{q}}{2}\Big(\sum_{a}J^a\Big)^2 -\frac{\hat{q}}{2}\sum_a(J^a)^2 \Big ) \\
&= \ln \int \frac{dt}{\sqrt{2\pi}} e^{-t^2/2} \left [ \int dJ \exp\left ( -\frac{E+\hat{q}}{2} J^2+(\hat{m}+\sqrt{\hat{q}}t) J\right )\right]^n.
\end{split}
\end{equation}
where we have applied the transformation to the third exponent in the first line. Using $a^n\approx 1+n\log a$ and $\log(1+a) \approx a$, we have
 \begin{equation}
W(\hat{m},\hat{q},E) = n \int \frac{dt}{\sqrt{2\pi}} e^{-t^2/2} \ln \left [ \int dJ \exp\left ( -\frac{E+\hat{q}}{2} J^2+(\hat{m}+\sqrt{\hat{q}}t) J\right )\right]
\label{Wrepsymm}
\end{equation}
In order to perform the Gaussian integrals one can show that for general $a$, $b$ parameters:
\begin{align*}
&\int dx e^{ax^2 \pm bx}=\sqrt{\frac{\pi}{a}}e^{\frac{b^2}{4a}}\\
&\int \frac{dx}{\sqrt{2\pi}}e^{-x^2/2}(a+b x)^2=a^2+b^2
\end{align*}
Therefore, integrating over $J$ in Eq. \eqref{Wrepsymm}, leads to: 
\begin{equation}
W(\hat{m},\hat{q},E) = n \Big ( \int \frac{dt}{\sqrt{2\pi}} e^{-t^2/2} \ln \sqrt{\frac{2\pi}{E+\hat{q}}} +   \int \frac{dt}{\sqrt{2\pi}} e^{-t^2/2} \frac{(\hat{m}+\sqrt{\hat{q}}t)^2}{2(E+\hat{q})} \Big )
\end{equation}
and integrating over $t$, finally leads to:
\begin{equation}
W(\hat{m},\hat{q},E)=\frac{n}{2}\Big [\ln(2\pi)-\ln(E+\hat{q})+\frac{\hat{q}+\hat{m}^2}{E+\hat{q}}\Big ]
\label{w_rsa}
\end{equation}
Computing $M$ is a bit more tricky. 
\begin{equation}
M(q,m)=\frac{p}{C}\ln \left [ \langle (1-\delta_{\eta^{\mu},0})I_1(q,m,\eta^{\mu}) +    \delta_{\eta^{\mu},0}I_2(q, m) \rangle_{\eta^{\mu}} \right ].
\end{equation}
as one has to compute $I_1(q,m,\eta^{\mu})$ and $I_2(q, m)$.
Using the Gaussian trick in Eq. \eqref{Gaussian} and assuming replica symmetry we rewrite Eq. \eqref{i1} as
\begin{equation}
\begin{split}
I_1(q,m,\xi)&= \int_{-\infty}^{\infty} \Big [ \prod_a \frac{dx_{\mu}^a}{2\pi}\Big] \exp \Bigg \{ \Big [ -\frac{i}{g}(\eta^{\mu}+g\vartheta)+i d_1^{inp} m \Big ] \sum_a x_{\mu}^a - \frac{d_3^{inp}}{2}\sum_a (x_{\mu}^a)^2+ -d_3^{inp} q \sum_{a<b} x_{\mu}^a x_{\mu}^b \Bigg \}\\
&= \int_{-\infty}^{\infty} \Big [ \prod_a \frac{dx_{\mu}^a}{2\pi}\Big] \exp \Bigg \{ \Big [ -\frac{i}{g}(\eta^{\mu}+g\vartheta)+i d_1^{inp} m \Big ] \sum_a x_{\mu}^a - \frac{d_3^{inp}}{2}\sum_a (x_{\mu}^a)^2+ \frac{d_3^{inp} q }{2} \sum_a (x_{\mu}^a)^2 - \frac{qd_3^{inp}}{2}\Big(\sum_a x_{\mu}^a \Big)^2\Bigg \}\\
&=\int Dt \Bigg \{ \int \frac{dx_{\mu}}{2\pi} \exp \Big [-i\left(g^{-1}\eta^{\mu} + \vartheta -d_1^{inp}m -t\sqrt{q d_3^{inp}}\right)x_{\mu} -\frac{d_3^{inp}}{2}(1-q)x_{\mu}^2 \Big ]  \Bigg \}^n
\end{split}
\label{i1_rsa}
\end{equation}
with $Dt=\frac{dt}{\sqrt{2\pi}}e^{-t^2/2}$. In a very similar way we can write Eq. \eqref{i2} as
\begin{equation}
I_2(q, m)=\int Dt \Bigg \{ \int_0^{\infty}\frac{d\lambda_{\mu}}{2\pi} \int_{-\infty}^{\infty}dy_{\mu} \exp \Big [ i \left(\lambda_{\mu}^a-\vartheta+d_1^{inp} m +t \sqrt{qd_3^{inp}}\right)y_{\mu} -\frac{d_3^{inp}}{2}(1-q)(y_{\mu})^2   \Big ] \Bigg \}^n.
\label{i2_rsa}
\end{equation}
We define $P(\eta^{\mu}>0)=f$
and rewrite Eq. \eqref{M} as
\begin{equation}
\begin{split}
M(q,m) &=\frac{p}{C}
\ln \{ 
\langle (1-\delta_{\eta^{\mu},0})\rangle_{\eta^{\mu}} 
\langle I_1(q,m,\eta^{\mu})\rangle_{\eta^{\mu}} 
+ 
\langle \delta_{\eta^{\mu},0}\rangle_{\eta^{\mu}} 
I_2(q, m) \}  \\
&=\frac{p}{C}\ln \Big [ f \langle I_1(q,m,\eta^{\mu})\rangle_{\eta^{\mu}} + (1-f) I_2(q, m) \Big ].
\end{split}
\label{M_rsaa}
\end{equation}
Simplifying for the sake of visualization Eq. \eqref{i1_rsa} and \eqref{i2_rsa} as
\begin{equation}
\begin{split}
&I_1 (q,m,\eta^{\mu}) = \int Dt Y^n\\
&I_2 (q,m) = \int Dt K^n
\end{split}
\end{equation}
where
\begin{equation}
\begin{split}
&Y\equiv \int \frac{dx_{\mu}}{2\pi} \exp \Big [-i\left(g^{-1}\eta^{\mu} + \vartheta -d_1^{inp}m -t\sqrt{q d_3^{inp}}\right)x_{\mu} -\frac{d_3^{inp}}{2}(1-q)x_{\mu}^2 \Big ] \\
& K \equiv \int_0^{\infty}\frac{d\lambda_{\mu}}{2\pi} \int_{-\infty}^{\infty}dy_{\mu} \exp \Big [ i \left(\lambda_{\mu}^a-\vartheta+d_1^{inp} m +t \sqrt{qd_3^{inp}}\right)y_{\mu} -\frac{d_3^{inp}}{2}(1-q)(y_{\mu})^2   \Big ]
\end{split}
\end{equation}
one can use again $a^n\approx 1+n\ln a$ and $\ln(1+a) \approx a$, which is valid for $n\rightarrow 0$, to write $M(q,m)$ as
\begin{equation}
\begin{split}
M(q,m)&=\frac{p}{C}\ln \Big [ f \left \langle \int Dt Y^n \right \rangle_{\eta^{\mu}} + (1-f)  \int Dt K^n\Big ]\\
& =\frac{p}{C}\ln  \Big [ \int Dt [f\left  \langle 1+n\ln Y\right  \rangle_{\eta^{\mu}} +(1-f)(1+n\ln K)\Big ]\\
&=\frac{p}{C}  \ln \Big [ 1+n \Big (  f \int Dt \left \langle \ln Y \right \rangle_{\eta^{\mu}} + (1-f) \int Dt \ln K \Big ) \Big ]\\
&= \frac{p}{C}n \Big (  f \int Dt \left \langle \ln Y \right \rangle_{\eta^{\mu}} + (1-f) \int Dt \ln K \Big )
\end{split}
\end{equation}
Turning back to the original notation we can further develop the terms composing the above approximation. The first one yields:
\begin{equation}
\begin{split}
\int Dt \left \langle \ln Y \right \rangle_{\eta^{\mu}}&= \int Dt \int \Bigg \langle \frac{dx_{\mu}}{2\pi} \exp \Big [-i\left (g^{-1}\eta^{\mu} + \vartheta -d_1^{inp}m -t\sqrt{q d_3^{inp}}\right )x_{\mu} -\frac{d_3^{inp}}{2}(1-q)x_{\mu}^2 \Big ] 	\Bigg \rangle_{\eta^{\mu}}		\\
&=\int Dt \Bigg \langle \ln \Big [ \exp \Big \{ -\frac{\left(d_1^{inp} m -g^{-1}\eta^{\mu}-\vartheta +t\sqrt{qd_3^{inp}}\right)^2}{2d_3^{inp}(1-q)}\Big \}\sqrt{\frac{2\pi}{d_3^{inp}(1-q)}}\frac{1}{2\pi}\Big ]\Bigg \rangle_{\eta^{\mu}}\\
&= \frac{1}{2}\left [ -\ln 2\pi -\ln d_3^{inp} (1-q) - \frac{\Big \langle \left(d_1^{inp} m -g^{-1}\eta^{\mu}-\vartheta\right)^2 \Big \rangle_{\eta^{\mu}}+qd_3^{inp}}{d_3^{inp} (1-q)}\right ]
\end{split}
\end{equation}
and the second one yields:
\begin{equation}
\begin{split}
\int Dt \ln K &= \int Dt \ln \int_0^{\infty}\frac{d\lambda_{\mu}}{2\pi} \int_{-\infty}^{\infty}dy_{\mu} \exp \left [ i \left (\lambda_{\mu}^a-\vartheta+d_1^{inp} m +t \sqrt{qd_3^{inp}}\right )y_{\mu} -\frac{d_3^{inp}}{2}(1-q)(y_{\mu})^2   \right ] \\
&=\int Dt \ln \int_0^{\infty}\frac{d\lambda_{\mu}}{2\pi}  \exp \left [ -\frac{\left (d_1^{inp}m +\lambda_{\mu -\vartheta + t\sqrt{qd_3^{inp}}}\right)^2}{1d_3^{inp}(1-q)}\right ] \sqrt{\frac{2\pi}{d_3^{inp}(1-q)}}\\
&= \int Dt \ln  \int ^{\infty}_{\frac{d_1^{inp}m -\vartheta+t\sqrt{qd_3^{inp}}}{\sqrt{d_3^{inp}(1-q)}}} \frac{dz}{\sqrt{2\pi}}e^{\frac{-z^2}{2}}
\end{split}
\end{equation}
where in the last passage we made a simple change of variables. Therefore we can rewrite Eq. \eqref{M_rsaa} as:
\begin{equation}
\begin{gathered}
M(q,m)=\frac{p}{C}n \Bigg \{
\frac{f}{2} \Bigg [ -\ln[2\pi d_3^{inp} (1-q)] - \frac{\left [\left \langle\left(d_1^{inp} m -g^{-1}  \eta^{\mu}  -\vartheta\right)^2\right \rangle_{\eta^{\mu}}+qd_3^{inp}\right]}{d_3^{inp}(1-q)}\Bigg ]
+(1-f)\int Dt \ln H(u) 
\Bigg \}\\
\text{where}\\
u\equiv \frac{d_1^{inp} m -\vartheta +t\sqrt{q d_3^{inp}}}{\sqrt{d_3^{inp} (1-q)}}\\
H(u)\equiv\int_u^{\infty}\frac{dz}{\sqrt{2\pi}}e^{-z^2/2}.
\end{gathered}
\label{M_rsa}
\end{equation}
Now we can evaluate the derivatives
\begin{equation}
\frac{dG}{d\hat{m}}=\frac{dG}{d\hat{q}}=\frac{dG}{dE}=\frac{dG}{dm}=\frac{dG}{dq}=0
\end{equation} 
where $G=G(q,\hat{q},m,\hat{m},E)$ given by Eq. \eqref{g_rsa}, and set them to zero to find the maximum of Eq. \eqref{g_rsa}, with $W(\hat{m},\hat{q},E)$ given by Eq. \eqref{w_rsa} and $M(q,m)$ given by Eq. \eqref{M_rsa}. \\
With the first three derivatives equalized to zero, which are applied only to the second and third term of Eq. \eqref{g_rsa}, and assuming $Cq\gg m^2$ and $|C(1-2q)|\gg m^2$ as $C\rightarrow \infty$, we obtain the relations
\begin{equation}
\begin{split}
&\hat{m}=-\frac{m}{\sqrt{C}(q-1)}\\
&\hat{q}=\frac{q}{(1-q)^2}\\
&E=\frac{1-2q}{(q-1)^2}.
\end{split}
\end{equation}
Substituting them into Eq. \eqref{g_rsa} we have to perform the last two derivatives.\\
$\frac{dG}{dm}$ can be simply evaluated, applying the Leibniz integral rule $\frac{d}{dx}[\int_{a(x)}^{b(x)} f(x,t)dt]=f(x,b(x))\frac{d}{dx}b(x)-f(x,a(x))\frac{d}{dx}a(x) +\int^{b(x)}_{a(x)}\frac{d}{dx}f(x,t)dt$ based on which $\frac{d}{dt} H(u(m))=\frac{d}{dm}\int^{\infty}_{u(m)}\frac{dz}{\sqrt{2\pi}}e^{\frac{-z^2}{2}}=-\frac{1}{\sqrt{2\pi}}e^{-\frac{u(m)^2}{2}}\frac{d}{dm}u(m)$ yielding:
\begin{equation}
\frac{dG}{dm}=0=-fd_1^{inp}(d_1^{inp} m -g^{-1}\langle \eta^{\mu} \rangle  - \vartheta) - \frac{\sqrt{d_3^{inp}(1-q)}(1-f)d_1^{inp}}{\sqrt{2\pi}} \int Dt H(u)^{-1}e^{-u^2/2}
\label{dgdm}
\end{equation}
The derivative in $q$ is a bit more tricky: 
\begin{equation}
\begin{gathered}
\frac{dG}{dq}=0=\frac{dM}{dq}+\frac{nq}{2(1-q)^2}\\
-\frac{nq}{2(1-q)^2}=\frac{p}{C} n \Bigg \{ 
- \frac{f}{2} \Bigg [ \frac{\langle(d_1^{inp} m -g^{-1} \eta^{\mu}   -\vartheta)^2\rangle+qd_3^{inp}}{d_3^{inp}(1-q)^2}\Bigg ]-\frac{(1-f)}{\sqrt{2\pi}}\int Dt e^{-u^2/2}\Bigg[ \frac{t\sqrt{d_3^{inp}}+(d_1^{inp}m-\vartheta)\sqrt{q}}{2\sqrt{d_3^{inp}}(1-q)\sqrt{q(1-q)}} \Bigg  ] H(u)^{-1}
\Bigg \}
\end{gathered}
\label{intermediatedgdq}
\end{equation}
where we have used as before $\frac{d}{dq} H(u(q))=\frac{d}{dq}\int^{\infty}_{u(t)}\frac{dz}{\sqrt{2\pi}}e^{\frac{-z^2}{2}}=-\frac{1}{\sqrt{2\pi}}e^{-\frac{u(q)^2}{2}}\frac{d}{dq}u(q)$ but as a function of $q$. Now the term multiplied by $(1-f)$ should be integrated by parts, i.e $\int_a^bu(x)v'(x)=u(b)v(b)-u(a)v(a) - \int_a^b u'(x)v(x)dx$. Remembering that $Dt\equiv \frac{dt}{\sqrt{2\pi}}e^{-t^2/2}$ one indeed can see that
\begin{equation}
\frac{d}{dt}\Big [e^{-\frac{t^2}{2}}e^{-\frac{u^2}{2}}\Big]=  - \Big ( t+u\sqrt{\frac{q}{1-q}}\Big )\Big (e^{-\frac{t^2}{2}}e^{-\frac{u^2}{2}}\Big ) =-\Bigg ( \frac{t\sqrt{d_3^{inp}}+(d_1^{inp} m -\vartheta)\sqrt{q}}{\sqrt{d_3^{inp}}(1-q)} \Bigg ) \Big (e^{-\frac{t^2}{2}}e^{-\frac{u^2}{2}}\Big )
\end{equation}
so one can re-write the term multiplied by $(1-f)$ in \eqref{intermediatedgdq} as
\begin{equation}
\begin{gathered}
\frac{(1-f)}{\sqrt{2\pi}}\int Dt e^{-u^2/2}\Bigg[ \frac{t\sqrt{d_3^{inp}}+(d_1^{inp}m-\vartheta)\sqrt{q}}{2\sqrt{d_3^{inp}}(1-q)\sqrt{q(1-q)}} \Bigg  ] H(u)^{-1} = -\frac{(1-f)}{2\sqrt{2\pi}\sqrt{q(1-q)}}\int \frac{dt}{\sqrt{2\pi}}\frac{d}{dt}\Big [e^{-\frac{t^2}{2}}e^{-\frac{u^2}{2}}\Big]H(u)^{-1}=\\
=-\frac{(1-f)}{2\sqrt{2\pi}\sqrt{q(1-q)}}\Bigg \{ \frac{1}{\sqrt{2\pi}}e^{-\frac{t^2}{2}}e^{-\frac{u^2}{2}}H(u)^{-1}\Big |^{t=+\infty}_{t=-\infty} - \int Dt e^{-\frac{u^2}{2}}\frac{d}{dt}H(u)^{-1}  \Bigg \}=\\
=\frac{(1-f)}{2\sqrt{2\pi}\sqrt{q(1-q)}}\Bigg \{ -\int Dt e^{-\frac{u^2}{2}}(-)H(u)^{-2}(-)\frac{1}{\sqrt{2\pi}}e^{-\frac{u^2}{2}}\frac{\sqrt{qd_3^{inp}}}{\sqrt{d_3^{inp}(1-q)}}\Bigg \}
\label{finderiv}
\end{gathered}
\end{equation}
where in the last passage we used again the Leibniz integral rule  with the derivative in $t$. Substituting back Eq. \eqref{finderiv} in the second term of Eq. \eqref{intermediatedgdq} and canceling out the repeated terms enables to reach right away the simplified solution:
\begin{equation}
\frac{dG}{dq}=0=\frac{\alpha}{q} \Bigg \{ f \Big [ \frac{\langle(d_1^{inp} m -g^{-1} \eta^{\mu}   -\vartheta)^2\rangle+qd_3^{inp}}{d_3^{inp}}\Big ] + \frac{(1-f)(1-q)}{2\pi}\int Dt H(u)^{-2}e^{-u^2} \Bigg\}
\label{dgdq}
\end{equation}
where $\alpha \equiv p/C$ is the storage load.\\
As explained in the main text we take the limit $q\rightarrow 1$, in which the storage load $\alpha$ becomes the critical capacity $\alpha_c$. Note that in this limit:
\begin{equation}
\lim_{q\rightarrow 1} u = 
\begin{cases} 
\infty  & \text{if } t>\frac{\vartheta -d_1^{inp} m}{\sqrt{d_3^{inp}}} \\
-\infty & \text{if } t<\frac{\vartheta -d_1^{inp} m}{\sqrt{d_3^{inp}}}. \\
\end{cases}
\end{equation}
and
\begin{align*}
&\lim_{u\to - \infty}H(u)\approx 1\\
&\lim_{u\to \infty}H(u) \approx  \frac{1}{\sqrt{2\pi}u}e^{-u^2/2} (1-\frac{1}{u^2})= \frac{1}{\sqrt{2\pi}u}e^{-u^2/2}
\end{align*}  
where in the second approximation we have Taylor expanded $H(u)$ around $u=0$.\\
This enables to further simplify the above equations, as one can define the variable
\begin{equation}
x=\frac{\vartheta-d_1^{inp} m}{\sqrt{d_3^{inp}}}
\label{xfin}
\end{equation} 
which can be used to divide the integral into two components, i.e. 
\begin{equation}
\int  Dt H(u)^{-\kappa}e^{-\kappa\frac{u^2}{2}}=\cancelto{0}{\int^x_{- \infty}  Dt H(u)^{-\kappa}e^{-\kappa\frac{u^2}{2}}} + \int_x^{\infty}  Dt H(u)^{-\kappa}e^{-\kappa\frac{u^2}{2}}
\end{equation}
where $\kappa=1$ in Eq. \eqref{dgdm} and $\kappa=2$ in \eqref{dgdq}.\\

The simple application of the limit $q\rightarrow 1$ with the above approximations, substituting back $u$ as in Eq. \eqref{M_rsa} and the new variable $x$ as in Eq. \eqref{xfin} leads to the final set of equations for the storage capacity
\begin{equation}
\begin{cases}
f(x+\frac{d_1^{out}}{g\sqrt{d_3^{inp}}})=(1-f)\int_x^{\infty}Dt (t-x)\\
\frac{1}{\alpha_c}=f \Big [ x^2+\frac{d_2^{out}}{g^2 d_3^{inp}}+\frac{2xd_1^
{out}
}{g\sqrt{d_3^{inp}}} +1 \Big ] + (1-f)\int_x^{\infty}Dt(t-x)^2.
\end{cases}
\label{final_sol_sep}
\end{equation}
where $d^{out}_{1,2,3}$ are defined in the same way as $d^{inp}_{1,2,3}$ except that the averages are now over the output distribution $\eta$.
 
Going from the calculation reported above for the threshold-linear perceptron it is straightforward to calculate the optimal capacity of a network of threshold linear units. Considering the network defined as in Eq. (1) of the main text, the corresponding volume we need to calculate can be written as 

\begin{equation}
V_T=\frac{ \int \prod_{i,j,j \neq i} d J_{ij} \delta\left (\sum_{j, j \neq  i} J^2_{ij} - C\right)\prod_{i,\mu}\Big[  \left(1-\delta_{\eta^{\mu},0}\right)\delta\left(h_i^{\mu} - \vartheta-\frac{\eta^{\mu}}{g}\right)+\delta_{\eta^{\mu},0}\Theta\left(\vartheta-h_i^{\mu}\right) \Big] }{\int \prod_{i,j,j \neq i} d J_{ij}  \prod_i \delta \left(\sum_{j,j \neq i} J^2_{ij} - C\right)}
\end{equation}
Since $V_T$ can be written as the product of the individual volumes of the connection weights towards each unit, as
$V_T=\prod_i^NV_i$ and thus $\langle \text{ln}V_T \rangle_{\eta}=N\langle \text{ln}V_i \rangle_{\eta}$, we will essentially be dealing with individual perceptrons like the one we have just studied. Putting $d_1^{inp}= d_1^{out}=d_1$ and $\; d_2^{inp}= d_2^{out}=d_2$ and thus $d^{inp}_3= d^{out}_3=d_3$ for $\forall i$, we arrive to the equations presented in Eq. (3) of the main text.\\
As explained in the main text, we evaluate the maximal storage capacity in the limit $g\rightarrow \infty$, which is approached already for moderate values of $g$. Eq. (3) of the main text in the $g\rightarrow \infty$ limit reduces to:
\begin{equation}
\begin{cases}
0=fx-(1-f)\int_x^{\infty}Dt (t-x)\\
\frac{1}{\alpha_c}=f  (x^2+1) + (1-f)\int_x^{\infty}Dt(t-x)^2,
\end{cases}
\label{final_sol_ginfi}
\end{equation}
which provides the universal $\alpha_c^G$ bound for errorless retrieval, dependent only through $f$ on the distribution of the patterns.

%==================================================================================
\subsection{Derivation of the limits}
From Eq. (3) of the main text it is possible to evaluate the two limits of very sparse and non-sparse coding. 
First, a simple substitution at $f=1$ leads to
\begin{align}
&x=-\frac{d_1}{g\sqrt{d_3}}\\
&\alpha_c^{-1}=1+\frac{1}{g^2}.
\end{align}

The limit $f\rightarrow 0$ is a bit trickier. We first rearrange the first equation in Eq. (3) as

\begin{equation}
\frac{f}{1-f}=\frac{1}{(x+\frac{d_1}{g\sqrt{d_3}})}\int_x^{\infty}Dt(t-x)=\frac{1}{(x+\frac{d_1}{g\sqrt{d_3}})} \Big( \frac{e^{-\frac{x^2}{2}}}{\sqrt{2\pi}} -x\int_x^{\infty}Dt \Big) 
\label{folimits}
\end{equation}

As $f$ goes to zero, for the left hand side to be equal to the right hand side, we should have $x\rightarrow \infty$. We therefore use the expansion
\begin{equation*}
\int_x^{\infty}Dt= \frac{e^{-\frac{x^2}{2}}}{\sqrt{2\pi}}\Big[\frac{1}{x}-\frac{1}{x^3}+\mathcal{O}\Big(\frac{1}{x^5}\Big)\Big]
\end{equation*}
to write the right hand side of Eq.\ \eqref{folimits} as
\begin{equation}
\frac{f}{1-f}\approx \frac{ e^{ -\frac{x^2}{2}}}{\sqrt{2\pi}x^3}.
\label{fxequation1}
\end{equation}

We find a solution to Eq. \eqref{fxequation1} through the following iterative procedure. We first solve the leading term for $f
\to 0$ in $x\to \infty$ namely
\begin{equation*}
f\approx\frac{ e^{ -\frac{x^2}{2}}}{\sqrt{2\pi}}.
\label{fxequation}
\end{equation*}
yielding 
\begin{equation}
 x\approx \sqrt{2\ln \Big (\frac{1}{\sqrt{2\pi}f} \Big )}
 \label{xfirstorder}
\end{equation}
We then insert $x$ from Eq.\ \eqref{xfirstorder} into $\exp(-x^2/2)=\sqrt{2\pi} f x^3$ to obtain the logarithmic correction
\begin{align}
e^{-\frac{x^2}{2}}&\approx\sqrt{2\pi}fx^3\nonumber\\
x &\approx \sqrt{2\ln\Big( \frac{1}{\sqrt{2\pi}f x^3}\Big)}\nonumber\\
x &\approx \sqrt{ 2\ln\Big(\frac{1}{\sqrt{2\pi}f}\Big) \Big(1 - \frac{\ln x^3}{\ln\frac{1}{\sqrt{2\pi}f}} } \Big ) \nonumber\\
& \approx \sqrt{ 2\ln\Big(\frac{1}{\sqrt{2\pi}f}\Big)} \Bigg(1 - \frac{3}{4}\frac{ \ln \Big ( 2\ln (\frac{1}{\sqrt{2\pi}f}) \Big )}{\ln \frac{1}{\sqrt{2\pi}f}} \Bigg ).
 \label{xsolution}
\end{align}
where in the last passage we have used the Taylor expansion of the square $\sqrt{1-y}=1-\frac{y}{2} +\mathcal{O}(y^2)$ around $y=0$ as for $f\rightarrow 0$, $\frac{\ln x^3}{\ln\frac{1}{\sqrt{2\pi}f}} \rightarrow 0$.

We have tested numerically that the above expression Eq.\ \eqref{xsolution} for $x$ is indeed a solution to Eq.\ \eqref{folimits} for $f\to 0$.

We now proceed to evaluate $\alpha_c$ and we apply the same Taylor expansion as before
\begin{align*}
\alpha_c & =\Big\{  f[\langle(x+\frac{\xi_i}{g\sqrt{d_3}})\rangle^2+1] + (1-f)\int_x^{\infty} Dt (t-x)^2 \Big\}^{-1}\\
&=  \Big\{   f[\langle(x+\frac{\xi_i}{g\sqrt{d_3}})\rangle^2+1] + (1-f)\Big(-\frac{xe^{-\frac{x^2}{2}}}{\sqrt{2\pi}} + (1+x^2)\int_x^{\infty} Dt \Big) \Big\}^{-1}\\
& \approx  \Big\{fx^2-\frac{xe^{-\frac{x^2}{2}}}{\sqrt{2\pi}} +\frac{(1+x^2)}{\sqrt{2\pi}}e^{-\frac{x^2}{2}}\Big(\frac{1}{x}-\frac{1}{x^3}+\frac{3}{x^5}\Big)\Big\}^{-1}\\
&\approx \Big\{fx^2+\frac{e^{-\frac{x^2}{2}}}{\sqrt{2\pi}}\Big (-x + \frac{(1+x^2)(x^4-x^2+3)}{x^5} \Big )\Big\}^{-1}\\
&\approx\Big\{fx^2+\frac{e^{-\frac{x^2}{2}}}{\sqrt{2\pi}}\Big( \frac{2x^2+3}{x^5} \Big )\Big\}^{-1}=\Big\{fx^2+\sqrt{\frac{2}{\pi}}\frac{e^{-\frac{x^2}{2}}}{x^3}\Big\}^{-1}.
\end{align*}

To summarize, in the limit $f\rightarrow 0$ we obtain
\begin{equation}
\begin{cases}
x \approx \sqrt{ 2\ln\Big(\frac{1}{\sqrt{2\pi}f}\Big)} \Bigg(1 - \frac{3}{4}\frac{ \ln \Big ( 2\ln (\frac{1}{\sqrt{2\pi}f}) \Big )}{\ln \frac{1}{\sqrt{2\pi}f}} \Bigg ) \\
\alpha_c \approx \Big\{fx^2+\sqrt{\frac{2}{\pi}}\frac{e^{-\frac{x^2}{2}}}{x^3}\Big\}^{-1}.
\end{cases}
\label{limita0}
\end{equation}
Substituting $x$ in $\alpha_c$ to the leading order leads to Eq.\ (5) presented in the main text.

%=============================================================================================================
%=============================================================================================================

\subsection{Details of the TL perceptron training algorithm}
For the purpose of assessing whether the Gardner capacity for errorless retrieval can be reached with explicit training, we can decompose a network of, say, $N+1=10001$ units into $N+1$ independent threshold linear perceptrons. 
A threshold linear perceptron is just a 1-layer feedforward neural network with $N$ inputs and one output, the activity of which is given by a threshold-linear activation function.
\begin{equation}
[h]^+=\text{max}(0,h)
\end{equation}
The network is trained with $p$ patterns. One can then think of the input as a matrix $\bar{\xi}$ of dimension $[N\times p]$ and of the output as a vector $\vec{\eta}$ of dimension $[1 \times p]$.

The aim of the algorithm is to tune the weights such that all $p$ patterns can be memorized. In order to tune the weights we start from an initial connectivity vector $\vec{J}_0$ of dimension $[1 \times N]$ and estimate the output $\vec{\hat{\eta}}$ as:
\begin{equation}
\begin{split}
&\vec{h}=\vec{J}\bar{\xi} \\
&\vec{\hat{\eta}}=g[\vec{h}]^+ \label{hateta}
\end{split}
\end{equation}
where $g$ is the gain parameter. We then compare the output $\vec{\hat{\eta}}$ with the desired output $\vec{\eta}$ through 
the loss function
\begin{equation}
{\rm L}(\vec{\hat{\eta}})=\sum_{\mu=1}^p \frac{1}{2}(\hat{\eta}^{\mu} -\eta^{\mu} )^2.
\end{equation}
The TL perceptron algorithm can be seen as simply a stripped down version of \textit{backpropagation}, for a 1-layer network: the weights $\vec{J}$ are 
modified by gradient descent to minimize the loss during the steps $k=1..k^{\textit{MAX}}$ where $k^{\textit{MAX}}$ is the number of steps needed for the gradient descent in order to reach the minima $\frac{d{\rm L}(\vec{J}_k)}{d\vec{J}_k}=0$. If at the minima $\rm L(\vec{J}_{\textit{k}^{\textit{MAX}}})=0$ at least a set of weights exists for errorless retrieval at that $p$ value. The storage capacity $\alpha_c=\frac{p^{max}}{N}$ is evaluated by estimating $p^{max}$ as the highest $p$ value enabling to reach $\rm L(\vec{J}_{\textit{k}^{\textit{MAX}}})=0$.\\
Initializing the weights around zero facilitates reaching the minima.
The chain derivative that in general implements gradient descent in backpropagation, in this case reduces to
\begin{equation}
\vec{J}_{k+1} = \vec{J}_{k} + \gamma\frac{g}{p} (\vec{\eta}-\vec{\hat{\eta}})\Theta(\vec{\hat{\eta}})\bar{\xi}^{\; T}
\label{jkp1}
\end{equation}
where $\Theta(\vec{\hat{\eta}})$ is the Heaviside step function applied to all $N$ elements of $ \vec{\hat{\eta}}$ and where $\gamma$ is a learning rate. Note that the gain $g$, appearing as a multiplicative factor both in Eq. \eqref{jkp1} and Eq. \eqref{hateta} is performing a similar role as the learning rate, with which it can be tuned.

In the simulations presented in Fig. 2 of the main text, in order to obtain the results shown by red diamonds we have used $N=100$ units, $g=1$ and binary patterns. For each value $f$, we increase $p$ and check whether when the connectivity matrix stops changing, we have  ${\rm L}(\vec{\hat{\eta}})=0$. We take $p_{max}$ as the largest value of $p$ for which this is possible for at least a set of random initial weights. As for the learning rate, we use a decreasing scheduling, initially set to  $\gamma=0.2$. As the minimization progresses, some weights stop changing while the others keep changing and therefore when there are only a maximum of $5$ weights changing, we decrease the learning rate to $\gamma=0.02$, and finally in later iterations when the number of still varying weights reduces to $2$ we use $\gamma=0.002$. The initial condition of the weights are drawn from a Gaussian distribution of mean $\mu=0$ and $\sigma=10^{-2}$ and each ($f,p$) combination is tested at least from $20$ random initial weights, each for a random data sample. In Fig. 2 of the main text we restrict our analysis to $N=100$ and $f\geq 0.05$ for numerical limitations. Decreasing $f$ implies on one side increasing the number of patterns, thus making the process slower; on the other side it reduces the possibility of finding non-zero values for finite $N$. Increasing $N$ in order to find non-zero values requires increasing $p$ accordingly, making the process even slower.\\
According to the analytical calculations, the same $f$ dependence of the capacity found for binary patterns should also hold for other distributions. In Fig. 1 we therefore also show the numerical experiments for input patterns taken at random from the ternary distribution $P(\eta)=(1-f)\delta(\eta) + \frac{f}{2}\delta(1-\eta) + \frac{f}{2}\delta(2-\eta)$ (green crosses in Fig. 1); the same distribution is also used for the outputs. The numerical results are consistent with the analytical results.
%=============================================================================================================
%=============================================================================================================

\subsection{Recap of the calculation of the Hebbian capacity in TL networks}
\label{sec:Hebbian_recap}
In this section we provide a brief recap of the main ideas and analytical tools and results reported in \citep{Tre90b, Tre+91, Tre91} about the storage capacity of networks of threshold-linear units. In the most general case, one considers that the threshold-linear unit $i$ receives an input
\begin{equation}
h_i=\sum_j J^c_{ij} V_j + b\left (\sum_j V_j/N \right)+\sum_{\mu} s^{\mu} \frac{\eta^{\mu}_i}{\langle\eta \rangle_\eta}
\end{equation}
where the first term is the standard term coming from the activity of the other units through the synaptic weights $J^c$. The second term is supposed to provide a general feedback, perhaps through inhibitory neurons that are not explicitly modelled, and it only depends on the mean network activity via a function $b$. The last term is the strength of the input aligned with one or more stored patterns. To study self-sustained attractors, we set $s^{\mu}$ to zero (which implies also $\delta=0$ in the notation of \cite{Tre91}). The unit activities $V_i$ are subject to the threshold-linear activation function, and the weight matrix is structured by Hebbian learning as
\begin{equation}
J_{ij}^C=c_{ij}\frac{1}{C \langle \eta \rangle_{\eta}^2 }\sum_{\mu=1}^p (\eta_i^{\mu}-\langle\eta\rangle_{\eta})(\eta_j^{\mu}-\langle \eta \rangle_{\eta}).
\label{jijc}
\end{equation}
were $C$ is the number of connections per unit and $c_{ij}=1$ if there is a connections from unit $j$ to unit $i$, $c_{ij}=0$ otherwise. 

Calculations for the storage capacity were performed for different types of network: the fully connected \cite{Tre90b}, the highly diluted \cite{Tre91}, the directed \cite{Tre+91} ones, and then generalised to arbitrary dilution in \cite{Rou+06}. Here we focus on networks with extremely diluted connectivity \cite{Tre91}, namely when $\frac{C}{N}\rightarrow 0$ and $C, N \rightarrow \infty$ such that the synapses $J_{ij}$ and $J_{ji}$ can be considered independent, and also on fully connected networks, useful to grasp a deeper understanding of the error at retrieval and the nature of Hebbian learning.

The calculation of the storage capacity involves the definition of the {\rm overlap} order parameters 
\begin{equation}
\hat{x}^{\mu}=\frac{1}{N}\sum_{i=1}^N\Big ( \frac{\eta_i^{\mu}}{\langle\eta\rangle_{\eta}}	-1 \Big )\langle V_i \rangle
\label{sub_over}
\end{equation}
measuring the overlap between the stored patterns and the activity of units $V_i$, where $\langle \cdots \rangle$ (without subscripts) denotes thermal average. One assume without loss of generality that one of the patterns, let us say the first pattern, is to be retrieved, and one then assumes the existence of stable states of the system for which $\hat{x}^{1}$ is non-zero while $\hat{x}^{\mu_s}, \mu \neq 1$ are zero in the thermodynamic limit. This is done in mean-field theory, developed either by means of the replica trick or signal-to-noise analyses, which yield self-consistent equations for the overlaps and other order parameters that appear in mean-field theory. An important order parameter that appears in the mean-field theory of attractor neural networks  is the variance of the quenched noise \cite{Ami92book}: it comes from the contribution to the field acting on each unit from the correlation of the activity of the network and that of non-retrieved patterns, i.e. those different from the first pattern. This correlation albeit small for each individual non-retrieved pattern make a significant contribution when $p$ is comparable to $C$ and is what makes retrieval impossible for large $p$. It thus needs to be included for calculating the storage capacity. Following \cite{Tre90b,Tre91}, we denote this parameter as $\rho$. In the case of threshold-linear two other order parameters are important that measure the relative magnitude of the signal (the part of the input to units that makes the units have the correct activity for retrieval) to the quenched noise $\rho$. The first one
\begin{equation}
w=\frac{b(x)-\hat{x}^1-\vartheta}{T_0\rho}
\end{equation} 
is the signal of the background versus the noise due to memory loading; $T_0=\langle \eta^2\rangle_{\eta}/\langle \eta\rangle^2_{\eta}$ and $\vartheta$ is the threshold (Eq. (1) of the main text). The second signal-to-noise ratio
\begin{equation}
v=\frac{\hat{x}^1}{T_0 \rho}.
\end{equation}
is specific to the units that have to be active.\\
The self-consistent mean-field equations that emerge from the calculations can be written in terms of the following quantities
\begin{eqnarray}
&A_1(w,v)=\frac{a}{v(1-a)}\Big\langle\Big ( \frac{\eta}{\langle\eta\rangle}-1\Big ) (x\phi(x)+\sigma(x))\Big\rangle_{\eta}-\langle \phi (x) \rangle_{\eta} \label{eq:A_1}\\
&A_2(w,v)=\frac{a}{v(1-a)}\Big\langle\Big ( \frac{\eta}{\langle\eta\rangle}-1\Big ) (x\phi(x)+\sigma(x))\Big\rangle_{\eta}\label{eq:A_2}\\
&A_3(w,v)=\Big\langle (x^2+1)\phi(x) + x\sigma(x) \Big\rangle_{\eta}\label{eq:A_3}\\
&A_4(w,v)=\frac{1}{v}\Big\langle (x\phi(x)+\sigma(x))\Big\rangle_{\eta}\label{eq:A_4}
\end{eqnarray}
where
\begin{eqnarray}
& x\equiv w+v\frac{\eta}{\langle \eta \rangle}  \label{x_nna}\\
& \phi(x)\equiv\frac{[1+\text{erf}(\frac{x}{\sqrt{2}})]}{2}=\frac{\text{erfc}(\frac{-x}{\sqrt{2}})}{2}\label{eq:phi}\\
& \sigma(x)\equiv\frac{e^{-x^2/2}}{\sqrt{2\pi}} \label{eq:sigma}\
\end{eqnarray}
and where the sparsity parameter, $a$, which was mentioned in Eq.\ (4) of the main text, defined as
\begin{equation}
a\equiv \frac{\langle \eta^2 \rangle}{\langle \eta \rangle_{\eta}^2}\label{eq:sparsity},
\end{equation}
shows up as a crucial quantity. 

As far as the calculation of capacity is concerned, for the fully connected network, these equations must satisfy the conditions
\begin{eqnarray}
&E^{fc}_1(w,v)=0=A_1(w,v)^2 -\alpha A_3(w,v)\label{eq:E_1fc}\\
&E^{fc}_2(w,v)=0=A_1(w,v)\Big ( \frac{1}{gT_0} - A_2(w,v) \Big )-\alpha A_2(w,v)\label{eq:E_2fc}
\end{eqnarray}
and for the highly diluted network
\begin{eqnarray}
&E^{hd}_1(w,v)=0=A_2(w,v)^2 -\alpha A_3(w,v)\label{eq:E_1hd}\\
&E^{hd}_2(w,v)=0=A_2(w,v)-\frac{1}{gT_0}. \label{eq:E_2hd}
\end{eqnarray}  
In other words, the storage capacity $\alpha_c$ can be computed by finding the largest value of $\alpha$ for which equation $E_1$ can be satisfied, while equation $E_2$ can be used to extract the optimal value of $g$.. 

The value of other order parameters, e.g. $\rho$, or $\hat{x}^2$ for each value of $\alpha$ and any given choice of the distributions of of $\eta$ can also be calculated as $\rho=xA_2/vA_4$, and $\hat{x}^1=T_0 \rho v$.

%=============================================================================================================
%=============================================================================================================

\subsection{The mathematical forms of the binary, ternary, quaternary and exponential distributions used in the main text}\label{Sec:form_distribution}
As reported in the main text, we have compared capacity values using a binary, ternary, quaternary and an exponential distribution:
\begin{align}
&p(\eta)=(1-a)\delta(\eta)+a\delta(1-\eta)\label{binar}\\
&p(\eta)=\Big(1-\frac{9a}{5}\Big)\delta(\eta)+\frac{3a}{2}\delta\Big(\eta-\frac{1}{3}\Big)+\frac{3a}{10}\delta\Big(\eta-\frac{5}{3}\Big)\\
&p(\eta)=\Big(1-\frac{9a}{4}\Big)\delta(\eta)+\frac{3a}{2}\delta\Big(\eta-\frac{2}{9}\Big)+\frac{3a}{5}\delta\Big(\eta-\frac{5}{9}\Big)+\frac{3a}{20}\Big(\eta -\frac{20}{9}\Big)\\
&P(\eta)=(1-2a)\delta(\eta)+4a\exp(-2\eta)\label{exp}
\end{align}
One can see that all distributions are such that $\langle \eta \rangle = \int _0^{\infty} d\eta P(\eta) \eta =a$ and $\langle \eta^2 \rangle = \int _0^{\infty} d\eta P(\eta) \eta^2 =a$, so that $a$ coincides with the sparsity $\langle \eta \rangle^2/\langle \eta^2 \rangle$ of the network. The fraction of active units is thus related to $a$ as $f=a, 9a/5, 9a/4$, $2a$ respectively.\\
One can also easily see that
\begin{equation}
\begin{split}
&A^{\rm binary}_2(w,v)=\frac{a}{v}\Big [ -w\phi(w)-\sigma(w) + \Big ( w +\frac{v}{a}\Big ) \phi \Big ( w +\frac{v}{a}\Big ) +\sigma \Big ( w +\frac{v}{a}\Big )\Big ]\\
&A^{\rm binary}_3(w,v)=(1-a)[(w^2+1)\phi(w)+w\sigma(w)]+a\Big\{\Big [\Big ( w +\frac{v}{a}\Big )^2+1\Big]\phi \Big ( w +\frac{v}{a}\Big ) + \Big ( w +\frac{v}{a}\Big ) \sigma \Big ( w +\frac{v}{a}\Big )\Big \}
\end{split}
\end{equation}
and the same can be explicitly defined also for the ternary and quaternary distributions. For the exponential one, instead, we derive it analytically in the following section.

\begin{figure}[htb]
    \centering
    \includegraphics[scale=0.15]{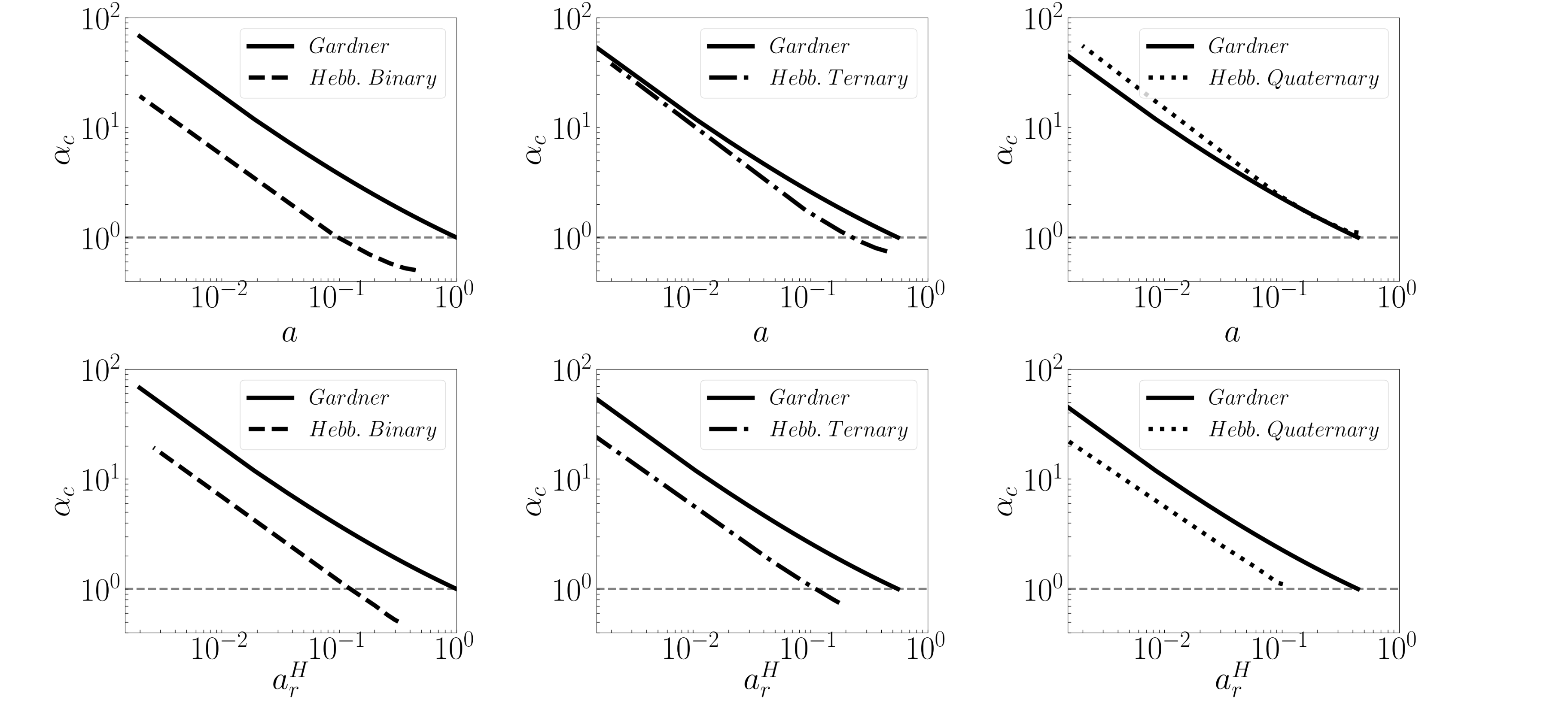}
    \caption{Suplementary to Fig. (2). Comparison between the Hebbian and Gardner storage capacity for 3 discrete distributions. The upper row considers as sparsity parameter the one of the input pattern, the lower row the one of the retrieved pattern. The Garner capacity is that given by Eq. (3) of the main text.}
   \label{6comparison_figure}
\end{figure}

As a supplement to Fig. 2 of the main text, reproduced here in the 3 separate panels in the upper row in Fig.\ \ref{6comparison_figure},  we show a comparison between the Hebbian capacity and the Gardner one when plotted as a function of the output sparsity (in the bottom row of Fig.\ \ref{6comparison_figure}). 
The Gardner storage capacity is now in each of these 3 cases above the Hebbian capacity, taken as a function of the output sparsity instead of the input one.

%=============================================================================================================
%=============================================================================================================

\subsection{Analytical derivation for the exponential distribution}
Here we report the explicit form of the expression for $A_2$ and $A_3$ in Eqs.\ \eqref{eq:A_2} and \eqref{eq:A_3} of Sect. \ref{sec:Hebbian_recap} for an exponential distribution of the patterns. In general, for $A_2$ we write
\begin{equation}
\begin{split}
A_2 &=\frac{a}{v(1-a)}\int_0^{\infty}d\eta P(\eta) (\frac{\eta}{\langle \eta \rangle}-1)\int_{-\infty}^{x(\eta)} Dz(x(\eta)-z) \\
&=\frac{a}{v(1-a)} \Big \{ \int_w^{\infty} Dz \int^{\infty}_{\frac{(z-w)\langle \eta \rangle}{v}} d\eta P(\eta) (\frac{\eta}{\langle \eta \rangle}-1)(x(\eta)-z) + \int_{-\infty}^{w} Dz \int^{\infty}_{0} d\eta P(\eta) (\frac{\eta}{\langle \eta \rangle}-1)(x(\eta)-z)  \Big \}
\end{split}
\end{equation}
with $x (\eta) \equiv w+v\eta / \langle \eta \rangle$.
Substituting Eq.\ \eqref{exp} we obtain
\begin{equation}
\begin{split}
& A^{\rm exp}_2=\frac{a}{v(1-a)} (A_{2.1}+A_{2.2}+A_{2.3} )\\
& A_{2.1}=\int_{-\infty}^w Dz\int_0^{\infty} d\eta 4a \exp(-2\eta)(\frac{\eta}{a}-1)(w+\frac{v\eta}{a}-z)\\
& A_{2.2}=\int_{-\infty}^w Dz (1-2a)(z-w)\\
& A_{2.3}=\int_w^{\infty} Dz \int_{\frac{(z-w)a}{v}}^{\infty}d\eta 4a \exp(-2\eta)(\frac{\eta}{a}-1)(w+\frac{v\eta}{a}-z).
\end{split}
\end{equation}
Solving the equations leads to
\begin{equation}
\begin{split}
&A_{2.1}=(1-2a)\sigma(w) + \Big [ \frac{v}{a} +w-v-2wa \Big ] \phi(w) \\
&A_{2.2}=(2a-1)(\sigma(w) +w\phi(w)) \\
&A_{2.3}=\exp \Big (\frac{2aw}{v} \Big )\exp \Big (\frac{2a^2}{v^2} \Big )\Big[ \frac{v(1-a)}{a}\phi \Big ( -w -\frac{2a}{v} \Big ) + \sigma 	\Big (w+\frac{2a}{v} \Big)-\Big (w+\frac{2a}{v} \Big )\phi \Big (-w -\frac{2a}{v} \Big) \Big]. \\
\end{split}
\end{equation}
Thus
\begin{equation}
A^{\rm exp}_2=\phi(w) + \exp \Big (\frac{2aw}{v} + \frac{2a^2}{v^2} \Big ) \Big \{ \phi \Big ( -w -\frac{2a}{v} \Big ) + \frac{a}{v(1-a)}\Big [ \sigma 	\Big (w+\frac{2a}{v} \Big)-\Big (w+\frac{2a}{v} \Big )\phi \Big (-w -\frac{2a}{v} \Big) \Big ] \Big \} 
\end{equation}

For $A_3$ we have
\begin{equation}
\begin{split}
& A^{\rm exp}_3=A_{3.1}+A_{3.2}+A_{3.3} \\
& A_{3.1}=\int_{-\infty}^w Dz\int_0^{\infty} d\eta 4a \exp(-2\eta)(w+\frac{v\eta}{a}-z)^2\\
& A_{3.2}=\int_{-\infty}^w Dz (1-2a)(w-z)^2\\
& A_{3.3}=\int_w^{\infty} Dz \int_{\frac{(z-w)a}{v}}^{\infty}d\eta 4a \exp(-2\eta)(w+\frac{v\eta}{a}-z)^2\\
\end{split}
\end{equation}

Substituting Eq.\ \eqref{exp} we obtain
\begin{equation}
\begin{split}
&A_{3.1}=(1-2a)\sigma(w) + \Big [ \frac{v}{a} +w-v-2wa \Big ] \phi(w) \\
&A_{3.2}=(2a-1)(\sigma(w) +w\phi(w)) \\
&A_{3.3}=\exp \Big (\frac{2aw}{v} \Big )\exp \Big (\frac{2a^2}{v^2} \Big )\Big[ \frac{v(1-a)}{a}\phi \Big ( -w -\frac{2a}{v} \Big ) + \sigma 	\Big (w+\frac{2a}{v} \Big)-\Big (w+\frac{2a}{v} \Big )\phi \Big (-w -\frac{2a}{v} \Big) \Big] \\
\end{split}
\end{equation}
and solving the equations leads to
\begin{equation}
\begin{split}
&A_{3.1}=2a \Big [ \sigma(w)(w+\frac{v}{a}) + \phi(w) (1+w^2+\frac{vw}{a}+\frac{v^2}{2a^2} \Big ] \\
&A_{3.2}=(1-2a)[w \sigma(w) +(1+w^2)\phi(w))] \\
&A_{3.3}=\frac{v^2}{a} \exp \Big (\frac{2aw}{v} \Big) \exp \Big (\frac{2a^2}{v^2} \Big )\phi \Big ( -w -\frac{2a}{v} \Big ). \\
\end{split}
\end{equation}
Thus
\begin{equation}
A^{\rm exp}_3=2v(\sigma(w)+\phi(w))+ w\sigma (w) + (1+w^2)\phi(w)+\frac{v^2}{a}\phi(w) + \exp \Big (\frac{2aw}{v} + \frac{2a^2}{v^2} \Big ) \phi(-w-\frac{2a}{v}).
\end{equation}

%=============================================================================================================
%=============================================================================================================

\subsection{Hebbian capacity of TL networks storing log-normal distributed patterns}
In the main text of the paper, we studied the storage capacity of TL networks when the neural activity of the stored patterns are drawn from a number of distributions: binary, ternary, quaternary and exponential. In particular, we analysed the experimental data in relation to the exponential distribution. 
Several authors, e.g. Buzs\'aki and Mizuseki, in \cite{buzsaki2014log}, have observed that often neural activity distribution resemble a log-normal distribution of suitable mean and variance.
While not claiming to perform a comprehensive model selection, their study makes the important point that neural activity in many instances has a heavier tail than Gaussian, and is better modelled by a log-normal distribution. In this section, we therefore analyze the storage capacity of TL networks with Hebbian learning also for patterns whose activity follows a log-normal distribution. To be concrete, we assume that the patterns $\eta$ are drawn from the following distribution
\begin{equation}
P(\eta)d\eta=\frac{1}{\eta}\frac{e^{-\frac{(\ln(\eta)-\mu)^2}{2\kappa^2}}}{\kappa\sqrt{2\pi}}d\eta
\label{eq:log-normal}
\end{equation}
for which we have
\begin{subequations}
\begin{align}
&\langle \eta \rangle_{\eta} = e^{\mu +\frac{\kappa^2}{2}}\\
&\langle \eta^2 \rangle_{\eta}  =e^{2\kappa^2+2\mu},
\end{align}
\end{subequations}
where here and in what follows $\langle \cdots \rangle_{\eta}$ represents averaging with respect to the log-normal distribution in Eq.\ \ref{eq:log-normal}. The sparsity, as defined in Eq.\ 4 of the main text, then reads
\begin{equation}
a=\frac{\langle \eta \rangle_{\mu}^2}{\langle \eta^2 \rangle_{\mu}}=e^{-\kappa^2}
\end{equation}
and it only depends on $\kappa$ and not on $\mu$.

If we substitute $z=\frac{\ln(\eta)-\mu}{k}$, such that $\eta=e^{zk+\mu}$ we obtain 
\begin{equation}
P(e^{\kappa z +\mu})dz=\frac{e^{-\frac{z^2}{2}}}{\sqrt{2\pi}}dz.
\end{equation}
Using this, and the fact that 
\begin{equation}
\frac{\eta}{\langle\eta\rangle}=e^{\kappa z + \mu - \mu -\frac{\kappa^2}{2}}=e^{\kappa z - \frac{\kappa^2}{2}},
\end{equation}
we can evaluate the quantities $A_2$ and $A_3$ defined in Eqs.\ \eqref{eq:A_2} and \eqref{eq:A_3} of Sect.\ \ref{sec:Hebbian_recap} as
\begin{subequations}
\begin{align}
&A^{\rm ln-n}_2(w,v)=\frac{a}{v(1-a)}\int_{-\infty}^{\infty} Dz \Big ( e^{\kappa z - \frac{\kappa^2}{2}}-1\Big ) \Big [ (w+ve^{\kappa z - \frac{\kappa^2}{2}})\phi(w+ve^{\kappa z - \frac{\kappa^2}{2}})+\sigma(w+v e^{\kappa z - \frac{\kappa^2}{2}})\Big ]\\
&A^{\rm ln-n}_3(w,v)=\int_{-\infty}^{\infty}Dz [(w+ve^{\kappa z - \frac{\kappa^2}{2}})^2+1]\phi(w+ve^{\kappa z - \frac{\kappa^2}{2}})+(w+ve^{\kappa z - \frac{\kappa^2}{2}})\sigma(w+ve^{\kappa z - \frac{\kappa^2}{2}})
\end{align}
\end{subequations}
which can then be used to find the storage capacity, $\alpha_c$, as the value of $\alpha$ above which Eq.\ \eqref{eq:E_1hd} in Sect. D cannot be satisfied. 

\begin{figure}[htb]
\centering
\includegraphics[scale=0.25]{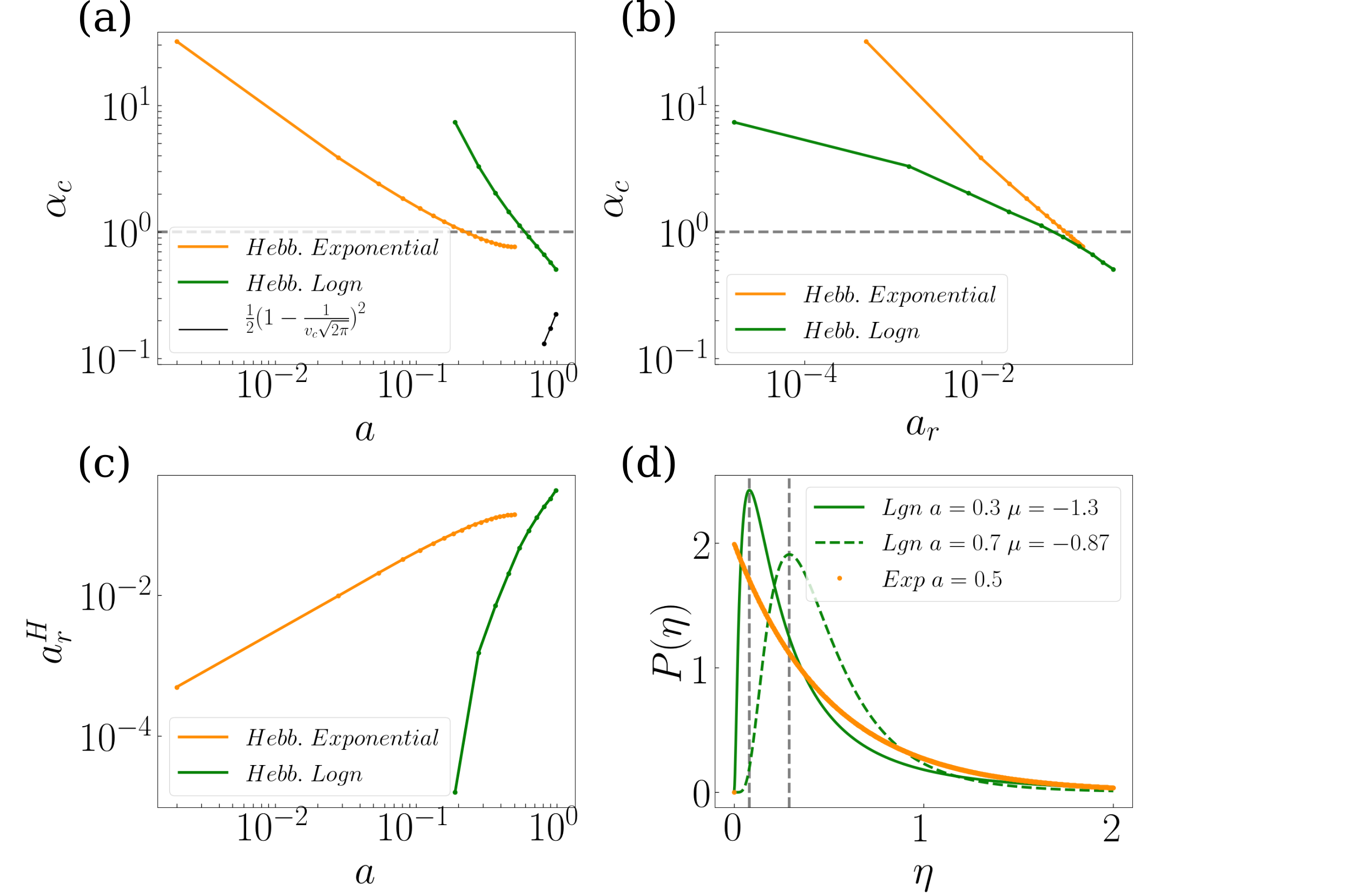}
\caption{a) $\alpha_c$ vs $a$, the black line corresponds to Eq. \eqref{lim} estimating the limit; b) $\alpha_c$ vs $a_r$; c) $a$ vs $a_r$. d) Examples of log-normal and Exponential distributions, with parameters (see the legend) such that $<eta>=0.5$ in all 3 cases. Note while the log-normal distributions have modes above zero, they have thinner tails than the exponential (but thicker than an ordinary normal). The vertical lines correspond to the maximum evaluated as in Eq. \eqref{eq: max_logn}}
\label{exp_logn}
\end{figure}

Fig.\ \ref{exp_logn}a shows $\alpha_c$ for the log-normal distribution as a function of the sparsity $a$. For comparison, we have also included the results for the exponential distribution. One can see that, when plotted as a function of the sparsity of the stored patterns, the capacity of the log-normal distributed patterns is higher than the exponential one. But this high capacity is obtained because of the much sparser retrieved pattern  compared to the stored one, as in Fig.\ \ref{exp_logn}c. When plotted versus the sparsity of the retrieved pattern, as can be seen Fig.\ \ref{exp_logn}b, the capacity of the log-normal distribution is always lower than the one of the exponential distribution. 
It is also possible to analytically find the limit of $a\to 1$ of the capacity of the log-normal distribution. 
In order to evaluate $\alpha_c=\frac{A_2^2}{A_3}$ as $a\to 1$ we estimate $A_2$ and $A_3$, by writing $b\equiv 1-a$, so that $b\to 0$. In this way: 
\begin{equation}
\begin{gathered}
k=\sqrt{-\ln(1-b)} \to \lim_{b\to0}k= \sqrt{b} + \mathcal{O} (b^{3/2})\\
\lim_{b\to 0}A_2=\frac{1-b}{vb}\int_{-\infty}^{\infty} Dz \Big ( e^{\kappa z - \frac{\kappa^2}{2}}-1\Big ) f(k)\\
f(k)=\Big [ (w+ve^{\kappa z - \frac{\kappa^2}{2}})\phi(w+ve^{\kappa z - \frac{\kappa^2}{2}})+\sigma(w+v e^{\kappa z - \frac{\kappa^2}{2}})\Big ]\\
\lim_{b\to 0}  e^{\kappa z - \frac{\kappa^2}{2}} = 1+\kappa z - \frac{\kappa^2}{2} +\mathcal{O}(k^3)
\end{gathered}
\end{equation}
so we have
\begin{equation}
\begin{gathered}
f(k)=f(k=0)+k\frac{f(k)}{dk}|_{k=0}+\mathcal{O}(k^3)\\
\frac{df(k)}{dk}=v\exp^{kz-\frac{k^2}{2}}(z-k)\phi(w+v\exp^{kz-\frac{k^2}{2}})\\
f(k)\approx (w+v)\phi(w+v)+\sigma(w+v) +kvz\phi(w+v)
\end{gathered}
\end{equation}
For $A_2(w,v)$ we have:
\begin{equation}
\begin{split}
\lim_{b\to 0}A_2&=\frac{1-b}{vb}\int_{-\infty}^{\infty} Dz \Big( \kappa z - \frac{\kappa^2}{2}\Big ) \Big [ f(k=0)+kvz\phi(w+v)\Big ]\\
&=\frac{1-b}{vb}\Big \{ k^2v\phi(w+v)\int_{-\infty}^{\infty} z^2 Dz -\frac{k^2}{2}f(k=0)\int_{-\infty}^{\infty} Dz \Big \} \\
&=\frac{1-b}{vb} k^2\Big (v\phi(w+v)-\frac{f(k=0)}{2}\Big)
\end{split}
\end{equation}
given that
\begin{equation}
k^2=-\ln(1-b) \approx b  +\mathcal{O}(b^2)
\end{equation}
then 
\begin{equation}
\lim_{b\to 0}A_2(w,v)=\lim_{a\to 1}A_2(w,v)=\phi(w,v)-\frac{(w+v)\phi(w+v)+\sigma(w+v) }{2v}
\end{equation}
For $A_3(w,v)$ instead we simply have
\begin{equation}
\lim_{a\to1} A_3(w,v)=[(w+v)^2+1]\phi(w+v)+(w+v)\sigma(w)
\end{equation}
Then
\begin{equation}
\lim_{a\to1}\alpha_c=\frac{\Big (\phi(w,v)-\frac{(w+v)\phi(w+v)+\sigma(w+v) }{2v}\Big)^2}{[(w+v)^2+1]\phi(w+v)+(w+v)\sigma(w)}
\end{equation}
Plotting $w_c+v_c$ as a function of $a$ one can see that $\lim_{a\to1} w_c+v_c\approx 0$.
If we substitute that we get
\begin{equation}
\lim_{a\to1} \alpha_c \approx \frac{1}{2}\Big( 1-\frac{1}{v_c\sqrt{2\pi}}\Big)^2 .
\label{lim}
\end{equation}
This equation was solved numerically up to the value of $a=0.99$, obtaining the black line in Fig.\ \ref{exp_logn}.

In order to estimate the exact value at $a\to1$ we can also require that the two derivatives vanish, i.e. $2A_2A_{2w}-\alpha A_{3,w}=0$ e $2A_2 A_{2v}-\alpha A_{3,v}=0$. To do so we define $x=w+v$ for simplicity of visualization and write
\begin{equation}
\begin{cases}
\left(\phi(x)-\frac{x\phi(x)+\sigma(x) }{2v}\right)^2-\alpha[x^2+1]\phi(x)+(x)\sigma(x)\\
2\left(\phi(x)-\frac{x\phi(x)+\sigma(x) }{2v}\right)\left(\sigma(x)+\frac{\phi(x)}{2v}\right)-2\alpha[x(\phi(x)+\sigma(x)]=0\\
2\left(\phi(x)-\frac{x\phi(x)+\sigma(x) }{2v}\right)\left(\sigma(x)+\frac{\phi(x)}{2v}-\frac{x\phi(x)+\sigma(x)}{2v^2}\right)-2\alpha[x(\phi(x)+\sigma(x)]=0
\end{cases}
\end{equation}
By subtracting the last two equations we obtain
\begin{equation}
2\left (\phi(x)-\frac{x\phi(x)+\sigma(x) }{2v}\right)\frac{x\phi(x)+\sigma(x)}{2v^2}=0
\end{equation}
which can be satisfied only if $v\to\infty$. One can then show that the rest of the equations hold for $x=0$, thus $\alpha_c^{logn}(a=1)=0.5$.\\
The maximum of the log-normal distribution defined in Eq. \eqref{eq:log-normal} is given by:
\begin{equation}
\frac{dP(\eta)}{d\eta}=\frac{P(\eta)}{\eta}\Big ( -1-\frac{\ln(\eta)-\mu}{k^2}\Big )=0
\end{equation}
where we get the condition $k^2-\ln(\eta) +\mu=0$ which is satisfied when:
\begin{equation}
\eta_{max}=ae^{\mu}
\label{eq: max_logn}
\end{equation}

In conclusion, activity distributions which are well fit by a log-normal result in associative networks that can operate in two somewhat distinct, but continuous regimes: if the distribution is tightly clustered around its mean, i.e. not sparse, $0.5 < a < 1$, the retrieved distribution is not sparse either, and the Hebbian capacity is between $\simeq 1$ and $0.5$, comparable but lower than the Gardner capacity. Note that for such values of $a$ no alternative exponential distribution is available, as it would imply $f=2a>1$, and indeed for the log-normal $f\equiv 1$ always (implying that a comparison with the Gardner bound would only be limited to its value for $f=1$, i.e. $\alpha^G(1)\equiv 1$). If instead the distribution is sparse, i.e. $k$ is larger such that $a<0.5$, the Hebbian capacity is above unity (but below that of the exponential fit, which has a fatter tail), but the retrieved distribution rapidly becomes so much sparser as to make retrieval unfeasible for any reasonably sized network.

%=============================================================================================================
%=============================================================================================================

\subsection{Calculating the sparsity of the retrieved patterns}
Following \citep{Tre+91}, the average of the activity and the average of the square activity in the patterns retrieved with Hebbian weights are calculated considering that the field, i.e. the input received by a cell with activity $\eta$ in the memory, is normally distributed around a mean field proportional to $x$. If we call $z$ a random variable normally distributed with mean zero and variance one, $x$ is already the mean field properly normalized. With the threshold-linear transfer function, the output will be $g(x+z)$ for $x+z > 0$ and $0$ with probability $\phi(-x)$.
Therefore the average activity $\langle V \rangle$ (denoted as $x$ in \cite{Tre90b, Tre91, Tre+91}) and the average square activity $\langle V^2 \rangle$ (denoted as $y_0$ in \cite{Tre90b, Tre91, Tre+91}) are,
\begin{subequations}
\begin{align}
& \left \langle V \right \rangle = g\left \langle \int_{-x(\eta)}^{\infty}Dz \left [x(\eta)+z\right ]\right \rangle_{\eta}=g\left \langle \left [x_c\phi(x_c) + \sigma(x_c)\right] \right \rangle_{\eta} \label{memmedia}\\
& \left \langle V^2  \right \rangle = g^2\left \langle \int_{-x(\eta)}^{\infty}Dz \left [x(\eta)+z\right ]^2 \right \rangle_{\eta}=g^2 \left \langle \left [(1+x_c^2)\phi(x_c) + x_c\sigma(x_c)\right]\right \rangle_{\eta},
\end{align}
\end{subequations}
where
\begin{equation}
x_c\equiv w_c+v_c\frac{\eta}{\langle \eta \rangle}.
\end{equation}
The sparsity of the retrieved memory is thus $a_r^H= \langle V \rangle^2/\langle V^2 \rangle$.

%=============================================================================================================
%=============================================================================================================

\subsection{Comparison with real data}
In the real activity distributions we use, each neuron emits, in time bins of fixed duration (we use 100msec), $0,\dots,n,\dots, n_{max}$ spikes, with relative frequency $c_n$, such that $\sum_{n=0}^{n_{max}} c_n=1$. These values are taken from Fig.\ 2 of \cite{Tre+99} and correspond to the histograms in Fig.2 below (and in Fig.3 of the main text); they are assumed to 
be the distributions of the patterns to be stored. If the weights are those described by the Gardner calculation, these patterns can be retrieved as they are, 
and their distribution remains the same. If they are stored with Hebbian weights close to the maximal Hebbian capacity, however, the retrieved distributions look different, and they can be derived as follows.

The firing rate $V$ of a neuron in retrieving a stored pattern $\eta$ is assumed proportional to $w+v\eta/\langle \eta \rangle+z$ \cite{Tre+91}, where the parameters $w$ and $v$ are appropriately rescaled signal-to-noise ratios (general and pattern-specific), such 
that the normally distributed random variable $z$, of zero mean and unitary variance, is taken to describe all other non constant (noise) terms, besides $\eta$ itself. Averaging over $z$ one can write, as in Eq.\ \eqref{memmedia}, that at the maximal capacity
\begin{equation}
    \left \langle V \right \rangle (\eta)= g \int_{-x(\eta)}^{\infty}Dz \left[x(\eta)+z \right]=g\left [x_c\phi(x_c) + \sigma(x_c)\right],
    \label{menomedia}
\end{equation}
where $ x(\eta) \equiv w+v\eta/\langle \eta \rangle $ and at the saddle-point the parameters $w$ and $v$ take the values $w_c$ and $v_c$ that maximize capacity, as explained in \citep{Tre+91}. This implies setting an optimal value for the threshold $\vartheta$, which in the analysis is absorbed into 
the parameter $w$, and which determines the sparsity of the retrieved distribution. The gain $g$ remains, however, a free parameter, that affects neither sparsity nor capacity. It is a rescaled version of the original gain $g$ in the hypothetical TL transfer function.
In other words, the maximal Hebbian capacity determines the shape of the retrieval activity distribution, but not its scale (e.g., in spikes per sec).

To produce a histogram, that details the frequency with which the neuron would produce $n$ spikes at retrieval, e.g. again in bins of 100msec, one has to set this undetermined scale. We set it arbitrarily, with the rough requirement that the frequency of producing $n_{max}$ spikes at retrieval  be below what it is in the observed distribution, 
taken to describe storage, and negligible for $n_{max}+1$ spikes. Having set the scale $g$, the frequency with which the neuron emits $n$ spikes at retrieval, with $0 < n < n_{max}$ is the probability that  $n-1/2 < V < n+1/2$, that is, it is a sum over contributions from each $\eta$, such that
\begin{equation}
\begin{split}
     n-\frac{1}{2} < g(w_c+v_c \frac{\eta}{\langle \eta \rangle}+z) < n+\frac{1}{2}  \\ 
      \frac{n}{g}-\frac{1}{2g} -x_c < z<   \frac{n}{g}+\frac{1}{2g} -x_c
    \label{probcn}
\end{split}
\end{equation}
i.e., 
\begin{equation}
     Pr(n) = \sum_{\eta=0}^{\eta_{max}} c_{\eta} \Big [\phi\Big ( \frac{n}{g}+\frac{1}{2g} -x_c\Big )- \phi\Big (\frac{n}{g}-\frac{1}{2g} -x_c \Big )\Big ], 
    \label{probccn}
\end{equation}
with appropriate expressions for the two extreme bins. These are the distributions shown in Fig.3 in the main text, and in Fig.2 below.

We took $g=\frac{1}{2}$, as this value satisfies the \textit{a priori} requirements and allows to keep the same number of bins in the retrieved memory as in the stored one (and the coefficients sum up to one, to a very good approximation).

%=============================================================================================================
%=============================================================================================================

\subsection{Analysis of the other recorded cells}
\begin{figure}[htb]
    \centering
    \includegraphics[scale=0.15]{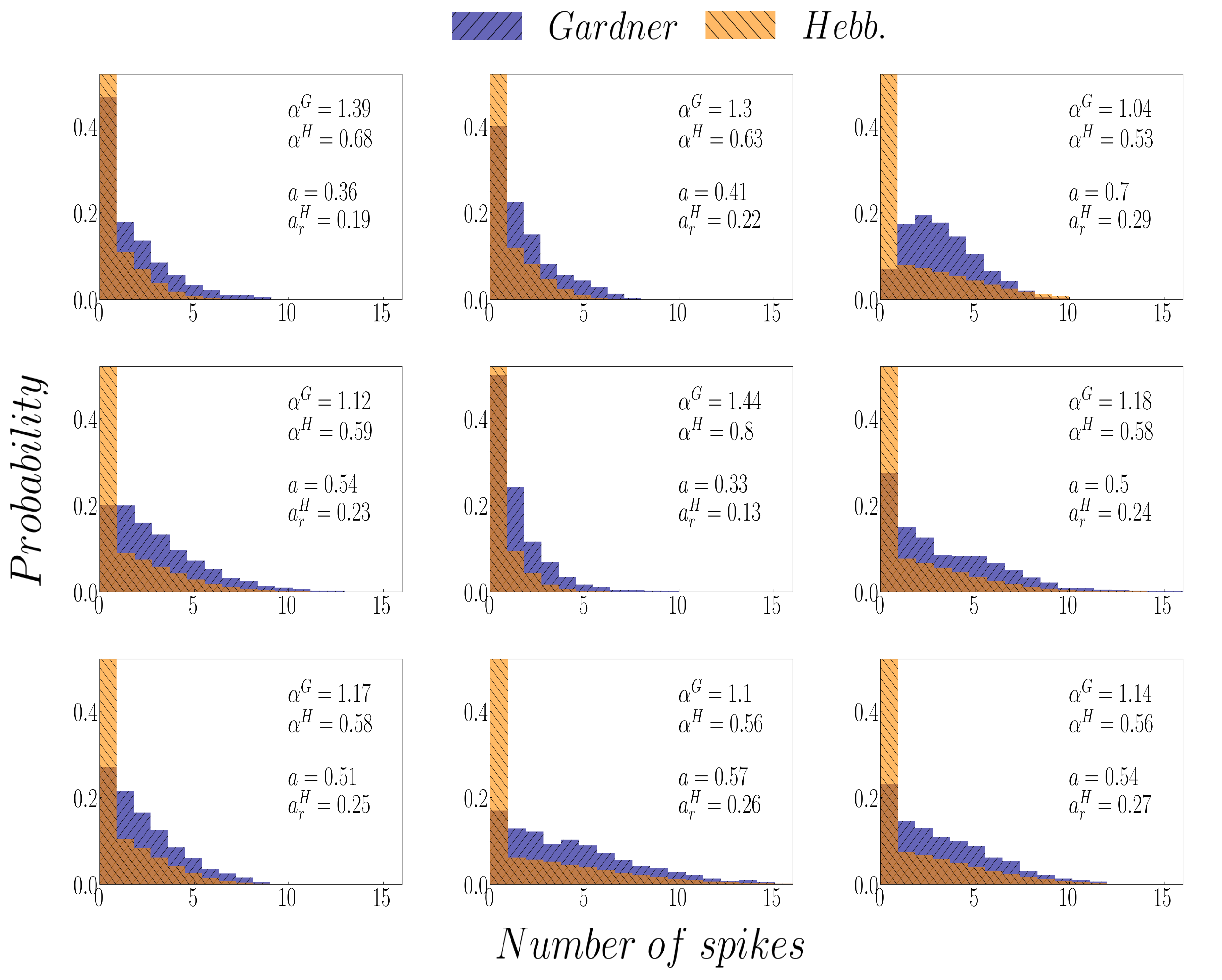}
    \caption{Suplementary to Fig (3) in the main text. }
   \label{supF3}
\end{figure}

Supplementary to Fig. (3) in the main text, we report in Fig. \ref{supF3} the same analysis for all $9$ single cells reported (using $100$ms bins) in \cite{Tre+99}. 

\begin{figure}[htb]
    \centering
    \includegraphics[scale=0.20]{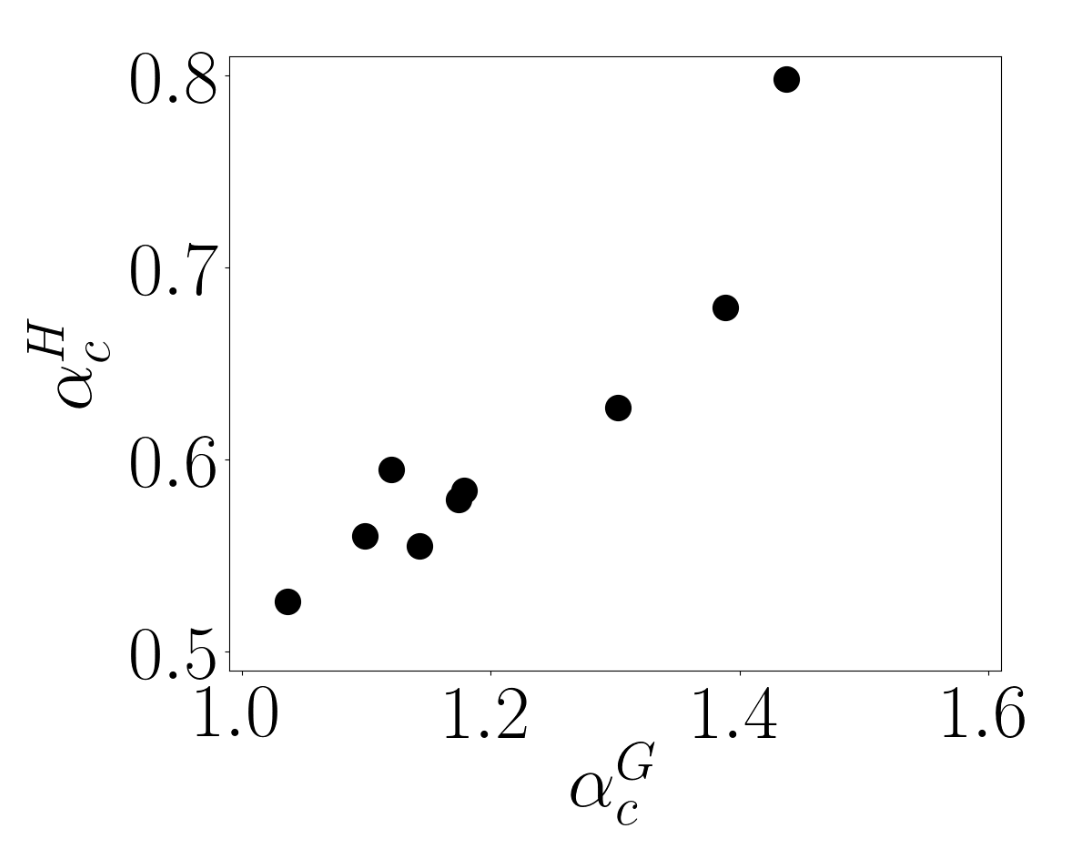}
    \caption{Comparison between the values of the storage capacity \textit{\`a la Gardner} and Hebbian, for the $9$ empirical distributions extracted from \cite{Tre+99}.}
   \label{alal}
\end{figure}

In each panel we write the capacity \textit{\`a la Gardner} and the Hebbian one (calculated without fitting an exponential) for the $9$ empirical distributions, as well as the sparsity of the original distribution and the sparsity of the one that would be retrieved with Hebbian weights. For simplicity of visualization we also show the storage capacity values against each other, calculated \textit{\`a la Gardner} and \textit{\`a la Hebb} (again, without fitting an exponential), as a single scatterplot  for the $9$ distributions, in Fig. \ref{alal}.
\bibliography{mybibAPS}{}